\documentclass[journal,draftcls,onecolumn,12pt]{IEEEtran}

\usepackage{cite}
\usepackage{graphicx}
\usepackage[cmex10]{amsmath}
\usepackage{amsfonts}
\usepackage{amssymb}

\ifCLASSOPTIONcompsoc
\usepackage[caption=false,font=normalsize,labelfont=sf,textfont=sf]{subfig}
\else
\usepackage[caption=false,font=footnotesize]{subfig}
\fi

\usepackage{amsthm}

\usepackage{algorithm}
\usepackage{algorithmic}

\usepackage{multirow}

\usepackage{tikz}
\usepackage{pgflibraryshapes}
\usetikzlibrary{arrows}
\usetikzlibrary{trees}
\usetikzlibrary{positioning}
\usetikzlibrary{plotmarks}
\tikzstyle{block} = [rectangle, very thick, draw=red!50!black!50, top color=white, bottom color=red!50!black!20, font=\itshape, text width=5.5em, text centered, rounded corners, minimum height=4em]
\tikzstyle{sblock} = [rectangle, very thick, draw=red!50!black!50, top color=white, bottom color=red!50!black!20, font=\itshape, text width=4em, text centered, rounded corners, minimum height=4em]
\tikzstyle{bgblock} = [very thick, draw=blue!50!black!50, bottom color=white, top color=blue!50!black!20, rounded corners, dashed]
\tikzstyle{bgblock2} = [very thick, draw=yellow!50!black!50, top color=white, bottom color=yellow!50!black!20, rounded corners, dashed]
\tikzstyle{bgblock3} = [very thick, draw=green!50!black!50, bottom color=white, top color=green!50!black!20, rounded corners, dashed]
\tikzstyle{arith} = [rectangle, minimum size=6mm, rounded corners=3mm, draw=black!50, very thick, top color=white, bottom color=black!20, text width=2em, text badly centered,  inner sep=0pt]
\tikzstyle{line} = [draw, thick, -latex', ->]

\pgfdeclarelayer{background}
\pgfdeclarelayer{foreground}
\pgfsetlayers{background,main,foreground}

\renewcommand{\theequation}{\arabic{section}.\arabic{equation}}
\newtheorem{remark}{Remark}

\newtheorem{result}{Result}
\newtheorem{theorem}{Theorem}
\newtheorem{proposition}{Proposition}
\newtheorem{lemma}{Lemma}
\newtheorem{corollary}{Corollary}

\theoremstyle{definition}

\theoremstyle{remark}

\newcommand{\defeq}{\mbox {$  \ \stackrel{\Delta}{=} $}}

\newcommand{\be}{ \mbox{$ \bf E $}}

\begin{document}

\title{New Non-asymptotic Random Channel Coding Theorems\thanks{This work was supported in part by the Natural Sciences and Engineering Research Council of Canada under Grant RGPIN203035-11, and by the Canada Research Chairs Program.}}

\author{En-hui Yang and Jin Meng\thanks{En-hui Yang and Jin Meng are with the Dept. of Electrical and Computer Engineering, University of Waterloo, Waterloo, Ontario N2L 3G1, Canada. Email: ehyang@uwaterloo.ca, j4meng@uwaterloo.ca} }

\date{ }
\maketitle

\newpage

\begin{abstract}
New non-asymptotic random coding theorems (with error probability $\epsilon$ and finite block length $n$) based on Gallager parity check ensemble and Shannon random code ensemble with a fixed codeword type are established for discrete input arbitrary output channels. The resulting non-asymptotic achievability bounds, when combined with non-asymptotic equipartition properties developed in the paper, can be easily computed. Analytically, these non-asymptotic achievability bounds are shown to be asymptotically tight up to the second order of the coding rate as $n$ goes to infinity with either constant or sub-exponentially decreasing $\epsilon$. Numerically, they are also compared favourably, for finite $n$ and $\epsilon$ of practical interest, with existing non-asymptotic achievability bounds in the literature in general.
\end{abstract}

\begin{IEEEkeywords}
Channel capacity, non-asymptotic coding theorems, non-asymptotic equipartition properties, random linear codes, Gallager parity check ensemble, Shannon random code, type.
\end{IEEEkeywords}

\newpage

\section{Introduction}
\label{sec:introduction}

Recently, there have been  great research interests in non-asymptotic channel coding theorems in information theory. By non-asymptotic coding theorems, we mean tight lower and upper bounds on the rate of certain codes or code ensembles in the regime of finite block length $n$ (typically ranging from hundreds to thousands) and (word) error probability $\epsilon$ (typically ranging from $10^{-1}$ to $10^{-9}$), which is loosely referred  to hereafter as  the non-asymptotic regime. For example, several non-asymptotic achievability bounds on Shannon random code ensemble have been reported in \cite{Yury-Poor-Verdu-2010}, which, coupled with non-asymptotic converse theorems therein, were shown to be very tight by numeric calculation in the non-asymptotic regime for some special channels such as a binary symmetric channel (BSC), a binary erasure channel (BEC),  and an additive white gaussian noise (AWGN) channel.

Following \cite{Yury-Poor-Verdu-2010}, we are motivated in this paper to investigate if similar tight bounds are still valid for some structured ensembles and general memoryless channels with finite input alphabet and arbitrary output alphabet. Of particular interest is Gallager parity check ensemble \cite{Gallager:LDPC1963}, in which  each element of the parity check matrix of a (linear) code is independently and uniformly generated from the finite field input alphabet. Note that for Gallager parity check ensemble, codewords are not pairwise independent, and therefore, bounding techniques on Shannon random code ensemble can not be applied in general.

Let $P = \{ p(y|x), x \in \mathcal{X}, y \in \mathcal{Y} \}$ be a channel with binary input alphabet $\mathcal{X}$. The channel $P$ is said to be memoryless binary-input output-symmetric (MBIOS) if  the transition probability distribution of the channel satisfies $p(y|0) = p(-y|1)$ for any $y \in \mathcal{Y}$. In the literature,  several non-asymptotic achievability bounds of linear codes have been developed for MBIOS channels. They more or less followed the approach invented by Gallager in \cite{Gallager:LDPC1963}. Specifically, given  a linear code $\mathcal{C}_n$ and a transmitted codeword $c^n$, the channel output space $\mathcal{Y}^n$ is divided into two parts $\mathcal{Y}^n_b$ (a bad region) and $\mathcal{Y}^n_g$ (a good region); the error probability (conditioned on the codeword $c^n$) then is bounded as follows
  \begin{eqnarray}
    \label{eq-gal-app}
    P_e (\mathcal{C}_n | c^n ) &\leq& \Pr \left\{ Y^n \in \mathcal{Y}^n_b | X^n = c^n \right\} \nonumber \\
                                &&{+}\: \Pr \left\{ \mbox{error},  Y^n \in \mathcal{Y}^n_g | X^n = c^n  \right\} ;
  \end{eqnarray}
and the union bound with respect to all codewords other than $c^n$ is then applied to the second probability term.  Using chernoff bounds\cite{chernoff}, Gallager \cite{Gallager:LDPC1963} then derived an achievability bound for any deterministic code of block length $n$ with respect to its Hamming weight profile $\{ N (l) \}^n_{i=1}$, where $N(l)$ is the number of codewords with Hamming weight $l$, and further showed that substituting $\{ N (l) \}^n_{i=1}$ in this achievability bound with the average Hamming weight profile of Gallager parity check ensemble yields a bound equal to the Error Exponent bound for Shannon random code ensemble in \cite{gallager:informtheory1968}, multiplied by a non-exponential term\footnote{This result on Gallager parity check ensemble was later enhanced by Shulman and Feder \cite{shu-feder1999}, who showed that the non-exponential term could be further eliminated.}. For some special MBIOS channels, analysis of those two probabilities in \eqref{eq-gal-app} can be further refined. Particularly, $\mathcal{Y}^n_b$ can be properly selected such that the exact calculation of the first probability is feasible for any finite block length, while for the second probability, the union bound can be applied conditioned on channel noise. Well known results along this line include those of  Poltyrev \cite{Poltyrev} for a BSC and binary input additive Gaussian channel (BIAGC). For BSCs, it was shown in \cite{Yury-Poor-Verdu-2010} that Poltyrev's bound on Gallager parity check ensemble turns out to be the tightest achievability bound in the non-asymptotic regime among all non-asymptotic achievabilities on BSCs in the literature. For BIAGCs, however, it was shown \cite{Poltyrev} that the corresponding bound (i.e., Tangential Sphere Bound (TSB)), applied to Gallager parity check ensemble, does not yield the same error exponent as that of Shannon random code ensemble (especially when the coding rate is close to Shannon capacity of the channel), and therefore would be expected to be worse than Error Exponent bound in the non-asymptotic regime. To the best of our  knowledge, for general MBIOS channels, Error Exponent bound remains the tightest achievability on Gallager parity check ensemble; it is also efficiently computable.



In this paper, a new non-asymptotic achievability bound is proved for Gallager parity check ensemble, which is applicable to any binary input memoryless channel\footnote{Our new  non-asymptotic achievability bound is also applicable to any memoryless channel with a finite field input alphabet. To facilitate our discussion, however, we choose to focus on the case of binary input alphabet when Gallager parity check ensemble is considered.} (BIMC). For some special channels such as BSCs and BECs, this bound can be calculated exactly, and is shown (both analytically and numerically) to be almost the same as Dependence Testing bound in \cite{Yury-Poor-Verdu-2010}. When combined with non-asymptotic equipartition property developed in the appendices of the paper, the new bound  can be efficiently evaluated for any BIMCs, including those with continuous output such as BIAGCs. Asymptotic analysis then shows that the new bound is tight up to the second order of the coding rate on any BIMC with certain symmetry  as $n$ goes to infinity with either constant or subexponentially
decreasing $\epsilon$. Numeric calculation on BIAGCs shows that the bound is tighter than TSB and Error Exponent bound in the non-asymptotic regime. Therefore, compared to Error Exponent bound, the tightest achievability bound (reported before in the literature) on Gallager parity check ensemble which is computable for general MBIOS channels, our achievability bound is more general (applicable to and computable for any BIMC with or without any symmetry) and tighter in the non-asymptotic regime.

Our bounding technique can be also applied to Shannon random code ensemble with a fixed codeword type on any discrete input memoryless channel (DIMC), in which each codeword is independently and uniformly generated from the set of sequences with the same type. The resulting achievability bound can be linked to $\kappa\beta$ bound, one of the tightest achievability bounds in the literature, proved in \cite{Yury-Poor-Verdu-2010} by a deterministically constructed code. Then an easy-to-compute version of the bound is yielded by applying non-asymptotic equipartition property, and is shown again to be tight up to the second order of the coding rate for any DIMC as $n$ goes to infinity with either constant or subexponentially decreasing $\epsilon$. Numerical calculation on Z channels shows that this achievability bound is tighter than Error Exponent bounds on Shannon random code with and without type constraint, derived by Fano \cite{fano1961} and Gallager \cite{gallager:informtheory1968} respectively,   in the non-asymptotic regime.

The rest of the paper is organized as follows. Non-asymptotic coding theorems for Gallager parity check ensemble on BIMCs and their asymptotic results are presented in Section \ref{sec4}, while their counterparts for Shannon random code ensemble with a fixed codeword type on DIMCs are presented in Section \ref{sec5}. Proofs of those theorems in Sections \ref{sec4} and \ref{sec5} are divided into Sections \ref{sec:proof-theorem-1}-\ref{sec:proof-theorem-3}. Section \ref{sec:comp-with-exist-achievable-bounds} is devoted to comparison between our non-asymptotic achievabilities and existing results in the literature, and the conclusion is drawn in Section \ref{sec8}.

\section{Non-asymptotic Coding Theorems for Gallager Parity Check Ensemble}
\label{sec4}
\setcounter{equation}{0}

In this section, we present non-asymptotic coding results for random linear codes of block length $n$ based on Gallager parity check ensemble for any BIMC. 

 Fix an arbitrary BIMC $\{p(y|x): x\in {\cal X}, y \in {\cal Y} \}$ with ${\cal X} = \{0, 1 \}$. Denote its channel capacity by $C_{\mathrm{BIMC}}$ and define its linear capacity as
\begin{displaymath}
  C_{\mathrm{BIMC-L}} = \ln 2 - H (X|Y)
\end{displaymath}
where $X$ is a uniform input random variable, and $Y$ is the corresponding output of the BIMC. (Here and throughout the rest of the paper, information quantities such as entropy, conditional entropy, mutual information, and divergence (or relative entropy) are measured in nats, and $\ln $ stands for the logarithm with base $e$.) Let $p(y)$ be the pmf or pdf (as the case may be) of $Y$, and $p(x|y)$ the conditional pmf of $X$ given by $Y$. It is easy to see that
\[ p(y) = {1 \over 2} [ p(y|0) + p(y|1) ] \]
and
\[ p(x |y) = {p(y|x) \over p(y|0) + p(y|1) }\;.\]

Let $\mathfrak{C}_{n,k}$ be a linear code with block length $n$ and parity check matrix $\mathbf{H}_{(n-k) \times n}$. Assuming codewords are ordered in some manner, we shall refer to the $q$-th codeword in $\mathfrak{C}_{n,k}$ as $x^n(q)$. We say $\mathbf{H}_{(n-k) \times n}$ is randomly picked from Gallager parity check ensemble $\mathcal{H}_{n,k}$ if entries of $\mathbf{H}_{(n-k) \times n}$ are independently and uniformly generated from ${\cal X} = \{ 0, 1 \}$. Denote the ensemble of linear codes with their parity check matrices from $\mathcal{H}_{n,k}$ by $\mathcal{C}^{(Gal)}_{n,k}$. To facilitate our subsequent discussion, we also specify the encoding procedure (i.e. the mapping from messages to codewords) of $\mathcal{C}^{(Gal)}_{n,k}$: given $\mathbf{H}_{(n-k) \times n}$, $x^n(q)$ is the $q$-th vector in the null space of $\mathbf{H}_{(n-k) \times n}$ by lexicographical order for $0 \leq q \leq 2^{n-rank(\mathbf{H}_{(n-k) \times n})}-1$. By convention, we assume that all messages are equally likely. With slight abuse of notation, we shall use $q$ to represent both the uniformly distributed random message and its specific realization; its exact meaning, however, will be clear from the context. Note that all codes in $\mathcal{C}^{(Gal)}_{n,k}$ have the channel coding rate greater than or equal to $\mathcal{R} (\mathcal{C}^{(Gal)}_{n,k}) \defeq \frac{k}{n} \ln 2$ (in nats). The decoding procedure (named as \textit{jar decoding}) is then specified as follows: given the channel output $y^n$, the decoder forms the set (also called BIMC-L jar for convenience)
\begin{equation} \label{eq2-bj}
  J(y^n) =\left  \{ x^n \in {\cal X}^n: - {1 \over n} \ln  { p(y^n
     |x^n) \over \prod_{i=1}^n [p(y_i |0) + p(y_i |1) ] } \leq  H(X|Y)
   + \delta \right \},
\end{equation}
declares an error if no codeword is inside $J(y^n)$, and pick an arbitrary codeword in $J(y^n)$ to be the estimate of the transmitted codeword otherwise. (Note that the case when more than one codeword is inside $J(y^n)$ is considered a tie by the decoder, which is broken in an arbitrary way\footnote{This decoding rule is closely related to Feinstein's threshold decoding. The difference lies in that when more than one codeword is inside jar or passes the threshold, the jar decoder treats the case as a tie, which is arbitrarily broken, while the threshold decoder will select the codeword with the lowest index. The reason for us to call this decoding rule jar decoding instead of modified threshold decoding is three fold: (1) it leads us to a philosophically different way to handle the second probability in \eqref{eq-gal-app}, as discussed in Remark \ref{reth2-0} and illustrated in the proof of Theorem \ref{thm-purerandom-bms-1}; (2) it allows us to easily identify which probability in \eqref{eq-gal-app} is dominating, as discussed in Remark \ref{rem-th-bimc}; and (3) by treating all codewords inside the jar equally, the decoder is not confined to solve any specific optimization problem, which, along with the flexibility of the formation of jar itself, we hope may lead one to look at practical decoding in a different way.}.)
It is easy to verify that
\begin{equation} \label{eq2-bjb}
  |J(y^n) | \leq e^{n ( H(X|Y) + \delta) }
\end{equation}
for any $y^n$.

Further define
\begin{equation}
  \label{eq-defn-pd}
  P_{\delta} \defeq  \Pr \left\{ - \frac{1}{n} \sum^n_{i=1} \ln p(X_i|Z_i) > H(X|Y) + \delta \right\}
\end{equation}
where $X_1 X_2 \cdots X_n$ is an independently, identically and uniformly distributed sequence and $Z_1Z_2 \cdots Z_n$ is the corresponding BIMC output.

Puncture $0$ from the message space and ignoring its insignificant effect on the rate, we have the following non-asymptotic coding theorem, which is proved in Section \ref{sec:proof-theorem-1}.
\begin{theorem}
  \label{thm-purerandom-bms-1}
  Given a BIMC with linear capacity $C_{\mathrm{BIMC-L}}$, let $P_e (\mathcal{C}^{(Gal)}_{n,k})$ denote the average word error probability (under jar decoding) of $\mathcal{C}^{(Gal)}_{n,k}$ with respect to the random message $q$, the BIMC, and the random linear code $\mathcal{C}^{(Gal)}_{n,k}$ itself. Then for any block length $n$ and $\delta > 0$
  \begin{equation}
    \label{eq-th-bms-1}
    P_e (\mathcal{C}^{(Gal)}_{n,k}) \leq \frac{1}{1-2^{-n}} P_{\delta} + e^{-n (C_{\mathrm{BIMC-L}} - \delta - \mathcal{R} (\mathcal{C}^{(Gal)}_{n,k}))} .
  \end{equation}
\end{theorem}

\begin{remark} \label{reth2-0} \em
  The key idea of the proof of Theorem \ref{thm-purerandom-bms-1}, as shown in Section \ref{sec:proof-theorem-1}, is to bound the error probability (under jar decoding) in two parts
  \begin{eqnarray*}
    P_e (\mathcal{C}^{(Gal)}_{n,k}) &\leq& \Pr \left\{ X^n (q) \notin J(Y^n) \right\} \\
                                             &&{+}\: \Pr \left\{ \exists z^n \in J(Y^n), z^n \neq X^n (q), z^n \in \mathcal{C}^{(Gal)}_{n,k}, X^n (q) \in J(Y^n) \right\} .
  \end{eqnarray*}
  Although this approach shares certain similarities with Gallager's proof technique illustrated in Section \ref{sec:introduction}, the key difference lies in that since all codewords inside the jar are treated equally, the second probability is handled by the union bound applied to all sequences inside $J(Y^n)$, instead of all codewords other than $X^n (q)$.  Therefore, no symmetry of channel is required in our proof.
\end{remark}

\begin{remark} \label{reth2-1} \em
The purpose of puncturing $q=0$ from the message space is to make the proof a little bit simpler. From the proof in Section \ref{sec:proof-theorem-1}, it can be seen that if we add $q=0$ back, it only increases the error probability upper bound by $2^{- n \mathcal{R}(\mathcal{C}^{(Gal)}_{n,k})}$. Moreover, when the channel has certain symmetry, i.e. $- \ln p(0|Y)$ given $X=0$ and $- \ln p (1|Y)$ given $X=1$ share the same distribution (we call such a channel  a binary input memoryless symmetric channel (BIMSC)), punctuation of zero message is not necessary and the term $\frac{1}{1-2^{-n}}$ in \eqref{eq-th-bms-1} can be dropped. 
 Note that the set of BIMSCs includes both MBIOS channels and weakly symmetric channels defined in \cite{cover:informtheory2006} as a special case, and in the case of BIMSC, $C_{\mathrm{BIMSC}} = C_{\mathrm{BIMSC-L}}$ always holds.
\end{remark}

\begin{remark} \label{reth2-1+} \em
The proof technique of Theorem \ref{thm-purerandom-bms-1} can be also applied to Shannon random code ensemble (with uniform input distribution) and Elias generator ensemble \cite{mct}, in which the generator matrices of linear codes are generated in the same way as that for parity check matrices in Gallager ensemble. In fact, the proof for those ensembles will be even simpler, 
and the term $\frac{1}{1-2^{-n}}$ in \eqref{eq-th-bms-1} can be dropped.
\end{remark}


As can be seen, the error probability bound in \eqref{eq-th-bms-1} is in a parametric form with respect to $\delta$. In other words, given the block length $n$ and the channel coding rate $\mathcal{R} (\mathcal{C}^{(Gal)}_{n,k})$ (or equivalently $k$), \eqref{eq-th-bms-1} holds for any value of $\delta$. And it is not hard to see that $P_{\delta}$ and $e^{-n (C_{\mathrm{BIMC-L}} - \delta - \mathcal{R} (\mathcal{C}^{(Gal)}_{n,k}))}$ are respectively decreasing and increasing functions of $\delta$. Consequently, there is an optimal $\delta$ which minimizes \eqref{eq-th-bms-1}. For some special channels such as BSCs and BECs, $P_{\delta}$ can be efficiently calculated for any $\delta$, and therefore the optimization of \eqref{eq-th-bms-1} with respect to $\delta$ can be exactly solved. However, for other channels, especially those with continuous output (like BIAGCs), it is extremely difficult to directly evaluate $P_{\delta}$. To overcome this problem, tight upper and lower bounds on $P_{\delta}$ are established in Appendix~\ref{sec:non-aasympt-equip-cond}. By combining these bounds on  $P_{\delta}$ with Theorem \ref{thm-purerandom-bms-1}, we then derive an achievability bound of an analytic form. Towards this, some definitions are needed.

 Let us temporarily drop the assumption that $\mathcal{X}$ is discrete and adopt the convention that $\int dx$ is interpreted as  $\sum_{x \in \mathcal{X}}$ when $\mathcal{X}$ is discrete. Now given a random variable pair $(X,Y)$ with distribution $p(x,y)$, let
\begin{displaymath}
  \lambda^*(X|Y) \defeq \sup \left\{ \lambda \geq 0: \iint p(y) p^{-\lambda+1} (x|y) dx dy < \infty \right\}\;.
\end{displaymath}
Suppose that
\begin{equation} \label{eq4-1}
  \lambda^*(X|Y) >0.
\end{equation}
Define for any $\delta \geq 0$
\begin{displaymath}
  r_{X|Y} (\delta) \defeq \sup_{\lambda \geq 0}
  \left[ \lambda \left(
      H (X|Y) + \delta
    \right) - \ln \iint p(y) p^{-\lambda + 1} (x|y) dx dy
  \right]
\end{displaymath}
and for $\lambda \in [0,\lambda^*(X|Y))$
\begin{displaymath}
  f_{\lambda} (x,y) \defeq \frac{p^{-\lambda}(x|y)}{ \iint p(v)p^{-\lambda+1}(u|v) du dv }
\end{displaymath}
\begin{displaymath}
  \delta(\lambda) \defeq \iint p(x,y) f_{\lambda} (x,y) [- \ln p(x|y) ] dx dy - H (X|Y)
\end{displaymath}
\begin{displaymath}
  \sigma^2_{H} (X|Y,\lambda) \defeq \iint f_{\lambda} (x,y) p(y) p(x|y) |- \ln p(x|y) - (H(X|Y)+\delta(\lambda))|^2 dx dy
\end{displaymath}
\begin{displaymath}
  M_{H} (X|Y,\lambda) \defeq \iint f_{\lambda} (x,y) p(y) p(x|y) |- \ln p(x|y) - (H(X|Y)+\delta(\lambda))|^3 dx dy
\end{displaymath}
\begin{eqnarray}
    \label{eq-nep-2-}
    \lefteqn{\bar{\xi}_{H}(X|Y,\lambda,n) = \frac{2CM_{H}(X|Y,\lambda)}{\sqrt{n} \sigma^3_{H}(X|Y,\lambda)}  } \nonumber \\
    &&{+}\: e^{\frac{n \lambda^2 \sigma^2_{H}(X|Y,\lambda)}{2}}
    \left[ Q(\sqrt{n}  \lambda \sigma_{H}(X|Y,\lambda)) - Q(\rho^*+\sqrt{n}  \lambda \sigma_{H}(X|Y,\lambda))\right]
  \end{eqnarray}
   where
 \[   Q(s) = {1 \over \sqrt{2 \pi}} \int_{s}^{\infty} e^{-x^2 / 2} d x \]
   $Q(\rho^*) = \frac{CM_{H}(X|Y,\lambda)}{\sqrt{n} \sigma^3_{H}(X|Y,\lambda)}$, and  $0 < C < 0.4784$  is the universal constant in the Berry-Esseen central limit theorem \cite{iid-clt-constant}.
  Denote $\sigma^2_{H} (X|Y, 0)$ by $\sigma^2_{H} (X|Y)$ and $M_{H} (X|Y, 0)$ by $M_{H} (X|Y)$, and define
\[ \Delta^* (X|Y) \defeq  \lim_{\lambda \uparrow  \lambda^*(X|Y)} \delta (\lambda) \]
where the above limit exists as shown in Appendix \ref{sec:non-aasympt-equip-cond}.
Further assume that
\begin{equation}
  \label{eq4-1-}
 \sigma^2_{H} (X|Y) >0 \mbox{ and } M_{H} (X|Y) < \infty.
\end{equation}

Now let $X$ be the uniform input random variable to the BIMC, and $Y$ the corresponding output random variable of the BIMC. Combining Theorem \ref{thm-purerandom-bms-1} with non-asymptotic bounds on $P_{\delta}$ developed in Appendix \ref{sec:non-aasympt-equip-cond}, we then get the following result, which is proved in Section \ref{sec:proof-theorem-2}.

\begin{theorem}
  \label{thm-purerandom-bms}
   For any BIMC with $ \sigma^2_{H} (X|Y) >0 $,   $\lambda^*(X|Y) >0 $, and $M_{H} (X|Y) < \infty$ and any block length $n$,   the following hold:
  \begin{enumerate}
  \item For any  $\delta \in (0, \Delta^* (X|Y))$
  \begin{equation}
    \label{eq-ch4-bms-1}
    P_e (\mathcal{C}^{(Gal)}_{n,k}) \leq \left( \frac{1}{1-2^{-n}} + \lambda \right) \bar{\xi}_H (X|Y, \lambda,n) e^{- n r_{X|Y} (\delta) }
  \end{equation}
  whenever
  \begin{equation}
    \label{eq-ch4-bms-2}
    \mathcal{R}(\mathcal{C}^{(Gal)}_{n,k}) \leq C_{\mathrm{BIMC-L}} - \delta -
    r_{X|Y} (\delta) + \frac{ \ln \lambda  \bar{\xi}_H (X|Y, \lambda,n) }{n}
  \end{equation}
  where $\lambda = r'_{X|Y} (\delta)$.

  \item  For  any real number $c$
   \begin{equation}
    \label{eq-ch4-bms-5}
    P_e (\mathcal{C}^{(Gal)}_{n,k}) \leq
    \frac{1}{1-2^{-n}} Q \left( \frac{c}{ \sigma_{H} (X|Y) } \right)
    + \frac{1}{\sqrt{n}} \left( \frac{C M_{H} (X|Y)}{\sigma^3_{H} (X|Y)} + \frac{e^{- \frac{c^2}{2 \sigma^2_{H} (X|Y)} }}{\sqrt{2 \pi} \sigma_{H} (X|Y)} \right)
  \end{equation}
  whenever
  \begin{equation}
    \label{eq-ch4-bms-6}
    \mathcal{R} (\mathcal{C}^{(Gal)}_{n,k}) \leq C_{\mathrm{BIMC-L}} - \frac{c}{\sqrt{n}}
    - \frac{\ln n}{2 n} - \frac{\frac{c^2}{2 \sigma^2_{H} (X|Y)} + \left[  \ln \sqrt{2 \pi} \sigma_{H} (X|Y) \right] }{n}.
  \end{equation}
    \end{enumerate}
\end{theorem}

\begin{remark} \label{rem-th-bimc} \em
  As shown in the proof of Theorem \ref{thm-purerandom-bms} in Section \ref{sec:proof-theorem-2}, given the coding rate $\mathcal{R} (\mathcal{C}^{(Gal)}_{n,k})$, the optimal $\delta$ is yielded by making
  \[ e^{-n (C_{\mathrm{BIMC-L}} - \delta + \mathcal{R} (\mathcal{C}^{(Gal)}_{n,k}))} \approx \lambda P_{\delta}\]
  and
  \[ e^{-n (C_{\mathrm{BIMC-L}} - \delta + \mathcal{R} (\mathcal{C}^{(Gal)}_{n,k}))} \approx  \frac{1}{\sqrt{n}} P_{\delta}\]
  in part 1) and 2) of Theorem \ref{thm-purerandom-bms} respectively. In both cases,
  \[ P_{\delta} \gg e^{-n (C_{\mathrm{BIMC-L}} - \delta + \mathcal{R} (\mathcal{C}^{(Gal)}_{n,k}))} \]
  for the optimal $\delta$ when $\mathcal{R} (\mathcal{C}^{(Gal)}_{n,k})$ is close to $C_{\mathrm{BIMC-L}}$. On the contrary, in Gallager's error exponent analysis illustrated in the introduction section, $\mathcal{Y}^n_b$ was chosen such that the first and second probabilities share the same exponent, for the sake of the tightness of error exponent. This difference, coupled with the fact that non-asymptotic bounds on $P_{\delta}$ in Appendix \ref{sec:non-aasympt-equip-cond} is tighter than chernoff bound, explains why our achievability can be tighter than Error Exponent bound in the non-asymptotic regime. Another advantage of applying non-asymptotic bounds on $P_{\delta}$ is that we do not have to choose $J(Y^n)$ for the sake of easy computation of $P_{\delta}$, which explains why our achievability can be tighter than TSB on BIAGC.
\end{remark}

\begin{figure}[h]
  \centering
  \includegraphics[scale=0.5]{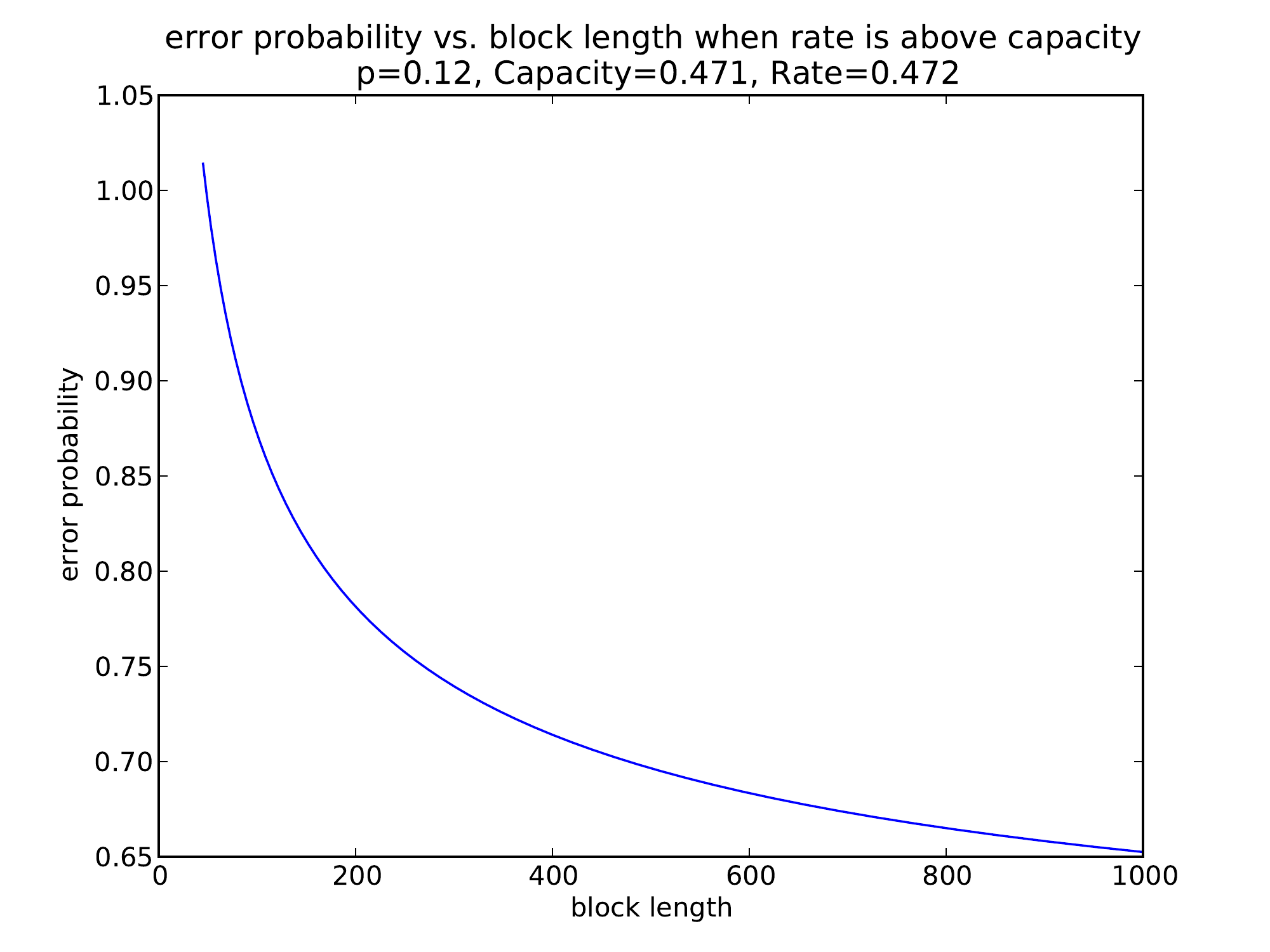}
  \caption{Tradeoff between the word error probability and block length when the code rate is above the capacity with $p=0.12$.}
  \label{abovecap}
\end{figure}

\begin{remark}  \label{reth2-3} \em
The inequalities \eqref{eq-ch4-bms-5} and \eqref{eq-ch4-bms-6} show that if the word error probability is kept slightly above $0.5$, the code rate can be even slightly above the capacity of the BIMC with $C_{\mathrm{BIMC}} = C_{\mathrm{BIMC-L}}$! Figure~\ref{abovecap} shows the tradeoff between the word error probability and block length when the code rate is $0.21\%$ above the capacity for the BSC with cross-over probability $p=0.12$, where in Figure~\ref{abovecap}, both the capacity and code rate are expressed in terms of bits. As can be seen from Figure~\ref{abovecap}, at the block length $1000$, the word error probability is around $0.65$, and the code rate is $0.21\%$ above the capacity! Although this phenomenon has been implied by the second order analysis of the coding rate as $n$ goes to $\infty$ \cite{strassen-1962,Yury-Poor-Verdu-2010,Hayashi-2009, allerton2012:channel, jar-converse} , the inequalities \eqref{eq-ch4-bms-5} and \eqref{eq-ch4-bms-6} allow us to demonstrate this for specific values of $n$ and for random linear codes based on Gallager parity check ensemble.
\end{remark}

\begin{remark} \label{rem-th-bimc-2} \em
  Parts 1) and 2) of Theorem \ref{thm-purerandom-bms} both provide non-asymptotic achievability bounds on the error probability and coding rate of Gallager's ensemble, which begs a comparison between them.
\begin{figure}[h!]
  \centering
  \includegraphics[scale=0.305]{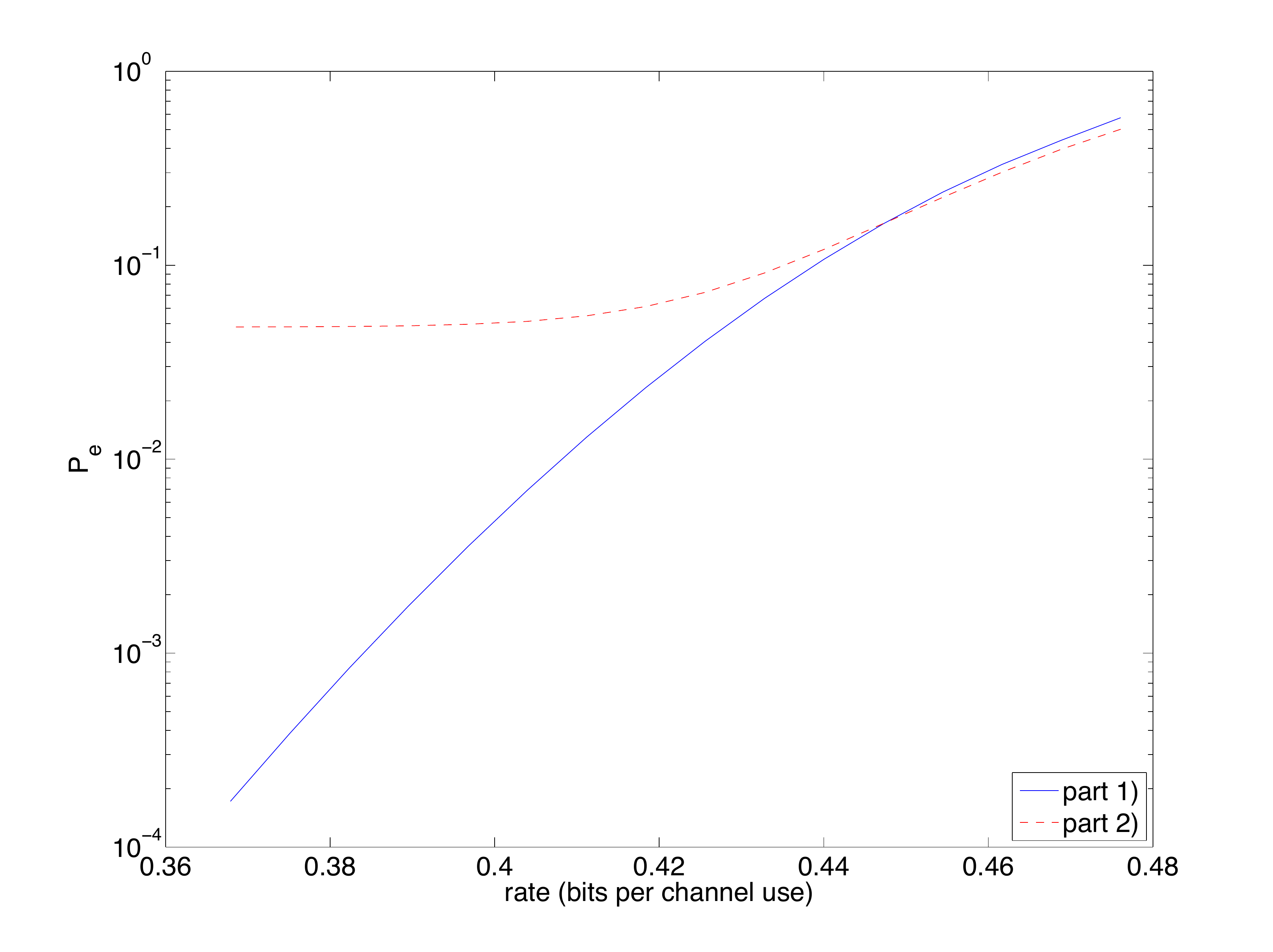}
  \caption{Part 1) vs Part 2) of Theorem \ref{thm-purerandom-bms} on BIAGC with block length $n=1000$ and snr=$0$dB}
  \label{part1vs2}
\end{figure}
It turns out that given block length, either of those achievability bounds can be tighter than the other for different coding rate regions. When the coding rate is above capacity, part 1) is not applicable, while part 2) can still bound the error probability strictly lower than $1$, shown in the above discussion. However, when the coding rate is below capacity, part 1) will be tighter than part 2) as long as the coding rate is not too close to the channel capacity. A numeric comparison between part 1) and part 2) is shown in Figure \ref{part1vs2} for BIAGC with block length $1000$ and snr 0dB, where the coding rate is kept less than the channel capacity $\approx 0.4847$ (bits per channel use). As can be seen, when the coding rate is moving away from the channel capacity, part 1) becomes much tighter.
\end{remark}

Although our focus in this paper is on non-asymptotic coding theorems, it is instructive to see how tight our achievability bounds in Theorem \ref{thm-purerandom-bms} are asymptotically as $n$ goes to $\infty$.
Then we get the following asymptotic result, which is proved in Section \ref{sec:proof-corollary-1}.

\begin{corollary}
  \label{cor-bimc-asym}
  Given a BIMC with $\sigma^2_{H} (X|Y) >0  $,   $\lambda^*(X|Y) >0 $, and $M_{H} (X|Y) < \infty$, let $\delta_n = \frac{\sigma_H (X|Y)}{\sqrt{n}} Q^{-1} (\epsilon_n)$ for  $0 < \epsilon_n <1$. Suppose  $\frac{- \ln \epsilon_n}{n} = o(1) $ as $n \rightarrow +\infty$. Then we have
  \begin{equation}
    \label{eq-cor-bimc-asym-1}
    \mathcal{R} (\mathcal{C}^{(Gal)}_{n,k}) \geq C_{\mathrm{BIMC-L}} - \delta_n - o \left( \delta_n \right)
  \end{equation}
  while $P_e (\mathcal{C}^{(Gal)}_{n,k}) \leq \epsilon_n$.
\end{corollary}

\begin{remark} \label{rem-bimc-asym} \em
  Given a BIMSC, results in \cite{strassen-1962,Yury-Poor-Verdu-2010,Hayashi-2009,moderate-deviation-arxiv,polyanskiy-allerton2010, allerton2012:channel, jar-converse} imply that $C_{\mathrm{BIMSC}} $ and $- \delta_n$ are the first and second order of the best coding rate that can be achieved by any code when the error probability is a constant or sub-exponentially decreasing with respect to $n$.  Corollary \ref{cor-bimc-asym} shows that the optimal first and second order coding performance can be achieved by Gallager ensemble under jar decoding as well. This in turn implies that the achievability bounds in Theorem \ref{thm-purerandom-bms} are asymptotically tight as $n$ goes to $\infty$ with either a constant or sub-exponentially decreasing  error probability with respect to $n$.
\end{remark}

\section{Non-asymptotic Coding Theorems for Shannon Random Code Ensemble with a Fixed Codeword Type}
\label{sec5}
\setcounter{equation}{0}

Consider now an arbitrary DIMC $P = \{p(y|x): x\in {\cal X}, y \in {\cal Y} \}$. Let $X$ be the capacity achieving input random variable. Let $Y$ be the output of the DIMC $P$ in response to $X$. Then the capacity of the DIMC $P$ is
 \[ C_{\mathrm{DIMC}} = I (X; Y) \;.\]

Now let us move away from linear codes in this section, and use random codes drawn from a particular type instead. Towards this, let us introduce some standard definitions involving types.  Let $\mathcal{P} (\mathcal{X})$ represent the set of all probability distributions on $\mathcal{X}$. For any $t \in \mathcal{P} (\mathcal{X})$, $t(x)$ denotes the probability of $x$ under $t$.  The set of types $\mathcal{P}_n (\mathcal{X})$ is the subset of $\mathcal{P} (\mathcal{X})$ such that $t \in \mathcal{P}_n (\mathcal{X})$ if and only if $t(x) n$ is an integer for any $x \in \mathcal{X}$. And for any $t \in \mathcal{P}_n(\mathcal{X})$, let $\mathcal{T}^n_{t} \subset {\cal X}^n $ be the set of sequences with empirical distribution $t$. Define for any $t \in \mathcal{P} (\mathcal{X})$
\begin{equation} \label{eq5-3-}
 D (t, x) \defeq \int p(y |x) \ln {p(y|x) \over q_t (y) } d y
  \end{equation}
  \begin{equation} \label{eq5-3}
  I ( t ; P ) \defeq \sum_{x \in \mathcal{X}} t(x) \int p (y|x)
  \ln \frac{p(y|x)}{q_t(y)} dy = \sum_{x \in \mathcal{X}} t(x) D(t, x)
  \end{equation}
where
\begin{displaymath}
  q_t (y) \defeq \sum_{x \in \mathcal{X}} t(x) p(y|x).
\end{displaymath}
Clearly, $D (t, x)$ is the divergence or relative entropy between $p(y|x)$ and $q_t (y)$;  and $I(t; P)$ is the mutual information between the input and output of the DIMC $ P$  when the input is distributed according to $t$. In addition, it can be easily verified that
\begin{equation} \label{eq5-4}
  I (t; P) = C_{\mathrm{DIMC}} + O(n^{-2})
\end{equation}
whenever
\begin{equation} \label{eq5-5}
  || t - p_X ||_1 \leq \frac{|\mathcal{X}|}{n}
\end{equation}
where $p_X$ is the capacity-achieving distribution, i.e. the distribution of $X$ maximizing $I(X;Y)$, and $||\cdot||_1$ is the $l_1$-norm. Obviously, types $t$ satisfying \eqref{eq5-5} exist.

Now let $\mathcal{C}_{t,n,k}$ denote the ensemble of channel codes from a type $t$ with code length $n$ and rate $\mathcal{R}(\mathcal{C}_{t,n,k}) = \frac{k}{n} \ln 2$, where a channel code from $\mathcal{C}_{t,n,k}$ is generated in such way that each codeword is independently and uniformly picked from $\mathcal{T}^n_{t}$. At the decoder, another version of jar decoding is used: given channel output $y^n$, the set $J(y^n)$ is formed as
\begin{equation} \label{eq5-6}
  J(y^n) = \left\{ x^n \in \mathcal{T}^n_{t}  :  - {1\over n} \sum^n_{i=1} \ln \frac{p(y_i|x_i)}{q_t (y_i)} < - I (t; P) + \delta \right\}
\end{equation}
where $\delta$ is a real number; then the decoder will declare an error if there is no codeword in $J(y^n)$ and pick an arbitrary codeword in $J(y^n)$ to be the estimate of the transmitted codeword otherwise. (Note that once again,  the case when more than one codeword is inside $J(y^n)$ is considered a tie, which is broken in an arbitrary way.)  The set defined in \eqref{eq5-6} will be referred to as the DIMC jar based on type $t$.

Define for any $x^n \in \mathcal{T}^n_{t}$
\begin{equation}
  \label{eq-defn-ptd}
  P_{t,\delta} = \Pr \left\{ \left. - {1\over n} \sum^n_{i=1} \ln \frac{p(Y_i|X_i)}{q_t (Y_i)} \geq - I (t; P) + \delta \right| X^n = x^n \right\}
\end{equation}
where $Y^n$ is the DIMC response to the input $X^n$. Note that $P_{t,\delta}$ is well defined since the probability on the right hand side of \eqref{eq-defn-ptd} depends on $x^n$ only through its type $t$. Then we have the following non-asymptotic coding theorem.
\begin{theorem}
  \label{thm-random-dimc-1}
  Given any DIMC $P$, let $ P_e (\mathcal{C}_{t, n,k})$  denote the average word error probability (under jar decoding) of $\mathcal{C}_{t, n,k}$ with respect to the DIMC and the random  code $ \mathcal{C}_{t, n,k}$ itself. Then for any block length $n$ and $\delta > 0$,
  \begin{equation}
    \label{eq-th-random-dimc-1}
    P_e (\mathcal{C}_{t,n,k}) \leq P_{t,\delta} + e^{- n( I(t;P) - \delta - \mathcal{R}(\mathcal{C}_{t,n,k})) + n H(t) - \ln |\mathcal{T}^n_t| } .
  \end{equation}
\end{theorem}

\begin{remark} \label{rem-thm-dimc} \em
  It is easy to show that
  \begin{equation}
    \label{eq-rem-thm-dimc-1}
    |\mathcal{T}^n_t| \geq \frac{1}{(n+1)^{|\mathcal{X}|}} e^{n H(t)}
  \end{equation}
  and therefore
  \begin{equation}
    \label{eq-rem-thm-dimc-2}
    n H(t) - \ln |\mathcal{T}^n_t| \leq |\mathcal{X}| \ln (n+1) .
  \end{equation}
  The term $ n H(t) - \ln |\mathcal{T}^n_t|$, instead of $ |\mathcal{X}| \ln (n+1)$,  is kept in \eqref{eq-th-random-dimc-1} to make the bound slightly tighter for small $n$.
\end{remark}

Similar to \eqref{eq-th-bms-1} in Theorem \ref{thm-purerandom-bms-1}, the achievability bound in \eqref{eq-th-random-dimc-1} in Theorem \ref{thm-random-dimc-1} holds for any $\delta > 0$ given the codeword type and the coding rate, and therefore the tightest bound is yielded by further optimizing $\delta$. When $P_{t,\delta}$ can not be efficiently calculated, an achievability bound of analytic form is needed. And once again, some definitions are in demand.

Given a DIMC $\{ p(y|x), x \in \mathcal{X}, y \in \mathcal{Y} \}$ and a distribution $t \in \mathcal{P} (\mathcal{X})$, let
 \begin{equation} \label{eq5-1}
  \lambda^*_{-} (t;P) \defeq \sup \left\{ \lambda \geq 0: \sum_{a \in \mathcal{X}} t(a) \int p(y|a)
    \left[ \frac{p(y|a)}{q_t(y)} \right]^{- \lambda} dy < +\infty \right\} .
\end{equation}
It is easy to see that $\lambda^*_{-} (t;P) $ depends on $t$ only through its support, i.e. $\{x \in \mathcal{X} : t(x) \neq 0 \}$. Suppose that
\begin{equation} \label{eq5-2}
  \lambda^*_{-} (t;P) >0 \;.
\end{equation}
Define any $\delta \geq 0$
\begin{displaymath}
  r_{-} (t,\delta) \defeq \sup_{\lambda \geq 0} \left\{ \lambda \left( \delta - I (t;P) \right) - \sum_{a \in \mathcal{X}} t(a) \ln \int p(y|a) \left( \frac{p(y|a)}{q_t(y)} \right)^{-\lambda} dy \right\}
\end{displaymath}
and for any $\lambda \in [0, \lambda^*_{-} (t; P))$
\begin{displaymath}
  f_{-\lambda,t} (y|x) \defeq \frac{\left[ \frac{p(y|x)}{q_t(y)} \right]^{-\lambda}}{ \int p(v|x) \left[ \frac{p(v|x)}{q_t(v)} \right]^{-\lambda} dv }
\end{displaymath}
\begin{displaymath}
  D (t,x,\lambda) \defeq \int p(y|x) f_{-\lambda, t} (y|x) \left[ \ln \frac{p(y|x)}{q_t(y)} \right] dy
\end{displaymath}
\begin{displaymath}
  \delta_{-} (t,\lambda) \defeq \sum_{x \in \mathcal{X}} t(x) \int p(y|x) f_{-\lambda, t} (y|x) \left[ - \ln \frac{p(y|x)}{q_t(y)} \right] dy + I (t; P) .
\end{displaymath}
Further define
\begin{displaymath}
  \sigma^2_{D,-} (t; P,\lambda) \defeq \sum_{x \in \mathcal{X}} t(x) \left[ \int p(y|x) f_{-\lambda,t} (y|x)  \left| \ln \frac{p(y|x)}{q_t(y)} - D (t,x,\lambda) \right|^2 dy \right]
\end{displaymath}
\begin{displaymath}
  M_{D,-} (t;P,\lambda) \defeq \sum_{x \in \mathcal{X}} t(x) \left[ \int p(y|x) f_{-\lambda, t} (y|x) \left| \ln \frac{p(y|x)}{q_t(y)} - D (t,x,\lambda) \right|^3 dy \right]
\end{displaymath}
and
\begin{eqnarray}
    \label{eqrl3-17-1-}
    \lefteqn{\bar{\xi}_{D,-}(t;P,\lambda,n) = \frac{2CM_{D,-}(t;P,\lambda)}{\sqrt{n} \sigma^3_{D,-}(t;P,\lambda)}  } \nonumber \\
                      &&{+}\: e^{\frac{n \lambda^2 \sigma^2_{D,-}(t;P,\lambda)}{2}}
                      \left[ Q(\sqrt{n}  \lambda \sigma_{D,-}(t;P,\lambda)) - Q(\rho^*+\sqrt{n}  \lambda \sigma_{D,-}(t;P,\lambda))\right]
  \end{eqnarray}
  with $Q(\rho^*) = \frac{CM_{D,-}(t;P,\lambda)}{\sqrt{n} \sigma^3_{D,-}(t;P,\lambda)}$.
Write $\sigma^2_{D, -} (t; P, 0)  $ simply as $\sigma^2_{D} (t; P)$, $M_{D, -} (t; P , 0)$ as $M_{D} (t; P )$, $\sigma^2_{D} (p_X; P)$ as $\sigma^2_{D} (X; Y)$, and $M_{D} (p_X; P )$ as $M_{D} (X; Y)$. It is not hard to see that
  \[ \sigma^2_{D} (t; P) = \sum_{x \in {\cal X}} t (x) \left [ \int p( y|x)  \left |
  \ln {p(y|x) \over q_t (y) }  \right |^2  d y - \left ( \int p( y|x)
  \ln {p(y|x) \over q_t (y) }    d y \right )^2  \right ] \]
and
  \[ M_{D} (t; P)  = \sum_{x \in {\cal X}} t (x) \left [ \int p( y|x)  \left |
  \ln {p(y|x) \over q_t (y) } -  \left ( \int p( v|x)     \ln {p(v|x) \over q_t (v) } dv \right )
 \right |^3  d y \right ] \;.\]
For obvious reasons, $\sigma^2_{D} (t; P)$ ($\sigma^2_{D} (X; Y)$, respectively) is referred to as the conditional divergence (or relative entropy\footnote{$\sigma^2_{D} (X; Y)$ coincides with channel dispersion defined in \cite{Yury-Poor-Verdu-2010}.}) variance of $P$ given $t$ ($Y$ given $X$, respectively).

Assume that
\begin{equation}
  \label{eq5-3+}
  \sigma^2_D (t;P) > 0 \mbox{ and } M_D (t;P) < +\infty.
\end{equation}
One can verify that Condition \eqref{eq5-3+} depends on $t$ only through its support; in other words, once Condition \eqref{eq5-3+} is valid for a distribution $t \in {\cal P}$, it is also valid for all distributions $\hat{t} \in {\cal P}$ with the same support as that of $t$.  In addition, it is not hard to verify that
 \[  \delta_{-} (t, 0)  =0 \]
 \begin{eqnarray*}
  {\partial  \delta_{-} (t, \lambda)  \over \partial \lambda}
  & = &
  \sum_{x \in {\cal X}} t (x) \left [ \int p( y|x) f_{-\lambda, t} (y |x) \left [- \ln {p(y|x) \over  q_t(y) } \right ]^2   dy   \right. \\
  & & \left.- \left ( \int p( y|x) f_{-\lambda, t} (y |x) \left [- \ln {p(y|x) \over  q_t(y) } \right ]   dy \right )^2  \right] \\
 & =&   \sum_{x \in {\cal X}} t (x) \left [ \int p( y|x) f_{-\lambda, t} (y |x) \left [ \ln {p(y|x) \over  q_t(y) } \right ]^2   dy   -  D^2 (t, x, \lambda) \right ] \\
 & > & 0
  \end{eqnarray*}
where the last inequality is due to \eqref{eq5-3+}. Therefore, $\delta_{-}(t, \lambda)$ as a function of $\lambda$ is strictly increasing over $\lambda \in [0, \lambda^*_{-} (t;P))$. Let
\[ \Delta^*_{-} (t) \defeq \lim_{\lambda \uparrow \lambda^*_{-} (t;P)} \delta_{-} (t, \lambda) \;.\]
It can be shown that $r_{ -} (t, \delta)$ is strictly increasing, convex and continuously differentiable up to at least the third  order inclusive over $\delta \in [0, \Delta^*_{-} (t))$, and furthermore $r_{-} (t, \delta)$ has the following parametric expression
\begin{equation} \label{eq3rp1}
  r_{ -} (t, \delta_{-}(t, \lambda)) =   \lambda ( \delta_{-} (t, \lambda) - I(t; P)) -
  \sum_{x \in {\cal X}} t(x) \ln  \int  p(y |x) \left [ {p (y|x) \over q_t (y) } \right ]^{-\lambda}  d y
\end{equation}
with
\[ \lambda = {\partial r_{-} (t, \delta) \over \partial \delta }\]
 satisfying
 \[ \delta_{-}(t, \lambda) = \delta \;.  \]

Then we get the following result, which can be proved in the same way as that for Theorem \ref{thm-purerandom-bms} (where non-asymptotic bounds on $P_{t,\delta}$ developed in Appendix \ref{sec:non-aasympt-equip-rela} are used), and therefore the proof of which is omitted.
\begin{theorem}
  \label{thm-random-dimc}
   For any DIMC $P$ and type $t$ satisfying \eqref{eq5-2} and \eqref{eq5-3+}, the following hold for any block length $n$:
  \begin{enumerate}
  \item For any $\delta \in (0, \Delta^*_{-} (t))$
  \begin{equation}
    \label{eq-thm-dimc-1}
    P_e (\mathcal{C}_{t, n, k}) \leq (1+ \lambda) \bar{\xi}_{D,-} (t;P,\lambda,n) e^{-n r_{-} (t,\delta)}
  \end{equation}
  whenever
  \begin{equation}
    \label{eq-thm-dimc-2}
    \mathcal{R} (\mathcal{C}_{t,n,k}) \leq I (t;P) - \delta - r_{-} (t,\delta) +
    \frac{\ln [ \lambda \bar{\xi}_{D,-} (t;P,\lambda,n)] - n H(t) + \ln |\mathcal{T}^n_t| }{n}
  \end{equation}
  where $ \lambda = {\partial r_{-} (t, \delta) \over \partial \delta }$ satisfying $\delta_{-} (t,\lambda) = \delta$.
  \item For any real number $c$
    \begin{equation}
      \label{eq-thm-dimc-5}
    P_e (\mathcal{C}_{t, n, k}) \leq Q \left(  \frac{c}{\sigma_{D} (t;P)} \right) + \frac{1}{\sqrt{n}} \left[ \frac{C M_{D} (t; P)}{\sigma^3_{D} (t; P)}  + \frac{e^{- \frac{c^2}{2 \sigma^2_D (t;P)} }}{\sqrt{2 \pi} \sigma_D (t;P)} \right]
    \end{equation}
    whenever
    \begin{equation}
      \label{eq-thm-dimc-6}
      \mathcal{R} (\mathcal{C}_{t,n,k}) \leq I(t;P) - \frac{c}{\sqrt{n}} - \frac{\ln n}{2 n} - \frac{\frac{c^2}{2 \sigma^2_D (t;P) } + \ln \left[ \sqrt{2 \pi} \sigma_D (t;P) \right] +n H(t) - \ln |\mathcal{T}^n_t| }{n}.
    \end{equation}
  \end{enumerate}
\end{theorem}

\begin{remark} \label{reth3-0} \em
  Comments similar to Remarks \ref{rem-th-bimc} to \ref{rem-th-bimc-2} immediately following Theorem \ref{thm-purerandom-bms} apply to Theorem \ref{thm-random-dimc} as well.
\end{remark}

\begin{remark} \label{reth3-1} \em
It is not hard to show that in the case of BIMC
\begin{equation} \label{eqth3and2}
   \sigma_{D} (X; Y) \leq \sigma_{H} (X|Y)
\end{equation}
and the inequality \eqref{eqth3and2} is strict in general unless the BIMC happens to be a BIMSC such as the BSC and BIAGC, in which case \eqref{eqth3and2} is the equality. Therefore, by comparing Theorem~\ref{thm-random-dimc} with Theorem~\ref{thm-purerandom-bms}, we see that for a BIMC which is not a BIMSC, Shannon random codes with a fixed codeword type are generally slightly better than random linear codes in terms of the tradeoff between the coding rate and word error probability. In addition, since our bounds in Theorem~\ref{thm-random-dimc} are valid for any $n$ and $t$, one can further optimize the bounds in Theorem~\ref{thm-random-dimc} over all input types satisfying \eqref{eq5-2} and \eqref{eq5-3+}.
\end{remark}

Given any DIMC $P$, fix a distribution $p_*$ on $\mathcal{X}$ satisfying \eqref{eq5-2} and \eqref{eq5-3+}. For any type $ t  \in \mathcal{P}_n (\mathcal{X})$ having the same support as that of $p_*$ and satisfying
  \begin{equation}
    \label{eq-asym-rem-1}
    \| t - p_* \|_1 \leq \frac{|\mathcal{X}|}{n}
  \end{equation}
 and for any $0< \epsilon_n  <1$,  let $\delta_{t, n} = \frac{\sigma_D (t;P)}{\sqrt{n}} Q^{-1} (\epsilon_n)$. In parallel with Corollary~\ref{cor-bimc-asym}, we have the following asymptotic result, which can be proved in a similar manner, and therefore the proof of which is omitted.

\begin{corollary}
  \label{cor-dimc-asym}
  Suppose $\frac{ \ln \epsilon_n}{n} = o(1) $ as $n \rightarrow +\infty$.  Then we have
  \begin{equation}
    \label{eq-asym-rem-2}
    \mathcal{R} ( \mathcal{C}_{t,n,k} ) \geq I(t;P) - \delta_{t,n} - o (\delta_{t,n})
  \end{equation}
  and
  \[ P_e (\mathcal{C}_{t,n,k}) \leq \epsilon_n \]
  for any type $ t  \in \mathcal{P}_n (\mathcal{X})$ having the same support as that of $p_*$ and satisfying \eqref{eq-asym-rem-1}.
  \end{corollary}

\begin{remark} \em
 In our companion paper\cite{jar-converse}, it is shown that $I(t;P)$ and $-\delta_{t,n} $ are the first and second order of the best coding rate that can be achieved by any code with its codewords drawn from ${\cal T}_t^n$ when the error probability is a constant or sub-exponentially decreasing with respect to $n$.  Corollary \ref{cor-dimc-asym} shows that the achievability bounds in Theorem \ref{thm-random-dimc} are asymptotically tight up to the second order as $n$ goes to $\infty$ with either a constant or sub-exponentially decreasing  error probability with respect to $n$.
\end{remark}

\section{Proof of Theorem \ref{thm-purerandom-bms-1}}
\label{sec:proof-theorem-1}
\setcounter{equation}{0}

Recall the encoding procedure of $\mathcal{C}^{(Gal)}_{n,k}$. Let $X^n (q)$ be the transmitted codeword, where $q$ is uniformly distributed over the punctured message space with message $0$ deleted. Let $Y^n$ be the output of the BIMC in response to $X^n (q)$. It is not hard to verify that for any $z^n \not = x^n \in {\cal X}^n$,
\begin{equation} \label{eqth2-2}
  \Pr \left\{ \left. z^n \in \mathcal{C}^{(Gal)}_{n,k}    \right | X^n (q) = x^n
         \right\} =   2^{-(n-k)} = e^{-(n-k) \ln 2}\;.
\end{equation}
To proceed, according to the decoding procedure specified in Section \ref{sec4}, we have
 \begin{eqnarray} \label{eqth2-3}
    P_e (\mathcal{C}^{(Gal)}_{n,k}) &\leq & \Pr \{ X^n (q) \not \in J(Y^n) \} \nonumber \\
    & & + \Pr \left \{  \exists z^n \neq X^n(q), z^n \in
        J(Y^n), z^n  \in \mathcal{C}^{(Gal)}_{n,k}, X^n (q)  \in J(Y^n)\right\} \nonumber \\
    & \leq & \Pr \{ X^n (q) \not \in J(Y^n) \}
    + \Pr \left \{  \exists z^n \neq X^n(q), z^n \in
        J(Y^n), z^n  \in \mathcal{C}^{(Gal)}_{n,k}\right\}
      \end{eqnarray}
where $J(Y^n)$ is the BIMC-L jar for $Y^n$. For any $x^n \in {\cal X}^n$ and $y^n \in {\cal Y}^n$, one can verify that
 \begin{eqnarray} \label{eqth2-4}
\lefteqn{ \Pr \left \{ \left.  \exists z^n \neq X^n(q), z^n \in
        J(Y^n), z^n  \in \mathcal{C}^{(Gal)}_{n,k}\right | X^n (q) = x^n, Y^n = y^n \right\}} \nonumber \\
   & = & \Pr \left \{ \left.  \exists z^n \neq x^n, z^n \in
        J(y^n), z^n  \in \mathcal{C}^{(Gal)}_{n,k}\right | X^n (q) = x^n, Y^n = y^n \right\} \nonumber \\
  & \stackrel{1)}{\leq} & \sum_{z^n \in J(y^n), z^n \not = x^n } \Pr \left \{ \left.   z^n  \in \mathcal{C}^{(Gal)}_{n,k}\right | X^n (q) = x^n \right\} \nonumber \\
   &\stackrel{2)}{\leq} & |J(y^n)| e^{-(n-k) \ln 2} \nonumber \\
   & \leq &     e^{n \left( H
        (X|Y) + \delta  \right)} e^{-(n-k) \ln 2}  =
    e^{- n \left( C_{\mathrm{BIMC-L}} - \delta - \mathcal{R}(\mathcal{C}^{Gal}_{n,k}) \right)
    }
  \end{eqnarray}
   where the inequality 1) follows from the fact that given $X^n (q)$, $Y^n$ and $\mathcal{C}^{(Gal)}_{n,k}$ are conditionally independent, the inequality 2) is due to \eqref{eqth2-2}, and finally the last inequality above is attributable to the upper bound on the size of the jar $J(y^n)$ in \eqref{eq2-bjb}. Since \eqref{eqth2-4} is valid for any $x^n \in {\cal X}^n$ and $y^n \in {\cal Y}^n$, it follows that
\begin{equation} \label{eqth2-5}
    \Pr \left \{  \exists z^n \neq X^n(q), z^n \in
        J(Y^n), z^n  \in \mathcal{C}^{(Gal)}_{n,k}\right\}
      \leq   e^{- n \left( C_{\mathrm{BIMC-L}} - \delta - \mathcal{R} (\mathcal{C}^{(Gal)}_{n,k} ) \right)
    } \;.
\end{equation}

To continue, let $X^n = X_1 X_2 \cdots X_n$ be a random variable taking values uniformly over ${\cal X}^n$. Let $Z^n =Z_1 Z_2 \cdots Z_n$ be the output of the BIMC in response to $X^n$. For $\mathcal{C}^{(Gal)}_{n,k}$, one can verify that  for any $x^n, x'^n \in \mathcal{X}^n / \{0^n\}$,
    \begin{eqnarray*}
    \Pr \left\{ X^n(q) = x^n \right\} &=& \sum_{\mathbf{H}_{(n-k) \times n}: \mathbf{H}_{(n-k) \times n} x^n = 0^{n-k}}
    \frac{2^{-(n-k)n} }{2^{ (n - rank ( \mathbf{H}_{(n-k) \times n}) )}-1} \nonumber \\
    &=& \sum_{\mathbf{H}_{(n-k) \times n}: \mathbf{H}_{(n-k) \times n} \mathbf{K}_{n \times n} x'^n = 0^{n-k}} \frac{2^{-(n-k)n} }{2^{ (n - rank ( \mathbf{H}_{(n-k) \times n}) )}-1} \nonumber \\
    &=& \sum_{\mathbf{H}_{(n-k) \times n}: \mathbf{H}_{(n-k) \times n} \mathbf{K}_{n \times n} x'^n = 0^{n-k}}\frac{2^{-(n-k)n} }{2^{ (n - rank ( \mathbf{H}_{(n-k) \times n} \mathbf{K}_{n \times n}) )}-1}\nonumber \\
    &=& \sum_{\mathbf{H}'_{(n-k) \times n}: \mathbf{H}'_{(n-k) \times n} x'^n = 0^{n-k}} \frac{2^{-(n-k)n} }{2^{ (n - rank ( \mathbf{H}'_{(n-k) \times n}) )}-1} \nonumber \\
    &=& \Pr \left\{ X^n(q) = x'^n \right\}
  \end{eqnarray*}
where $\mathbf{K}_{n \times n}$ is an invertible matrix such that $x^n = \mathbf{K}_{n \times n} x'^n$. This implies that for $\mathcal{C}^{(Gal)}_{n,k}$, $X^n (q)$ takes all sequences $x^n  \in \mathcal{X}^n / \{0^n\}$  equally likely. Since the zero sequence is not allowed by way of puncturing, it follows that the distribution of $X^n (q)$ is the same as the conditional distribution of $X^n$ given $X^n \not = 0^n$. Therefore, we have
\begin{eqnarray} \label{eqth2-7}
  \Pr \{ X^n (q) \not \in J(Y^n) \}  & =  & \Pr \{ \left. X^n  \not \in J(Z^n) \right | X^n \not = 0^n \}  \nonumber \\
  & \leq & { 1 \over 1 - 2^{-n} } \Pr \{ X^n  \not \in J(Z^n)  \} \;.
\end{eqnarray}

Putting \eqref{eqth2-3} and  \eqref{eqth2-5}-\eqref{eqth2-7} together yields
\begin{equation} \label{eqth2-8}
   P_e (\mathcal{C}^{(Gal)}_{n,k}) \leq \frac{1}{1-2^{-n}} \Pr \{ X^n  \not \in J(Z^n)  \} +
   e^{- n \left( C_{\mathrm{BIMC-L}} - \delta - \mathcal{R} (\mathcal{C}^{(Gal)}_{n,k} ) \right)}
\end{equation}
and the theorem is proved by observing that
\begin{equation}
  \label{eqth2-8+}
  \Pr \{ X^n  \not \in J(Z^n)  \} = \Pr \left\{ - \frac{1}{n} \sum^n_{i=1} \ln p(X_i|Z_i) > H(X|Y) + \delta \right\} = P_{\delta}
\end{equation}
due to the definition of the BIMC-L jar.

\section{Proof of Theorem \ref{thm-purerandom-bms}}
\label{sec:proof-theorem-2}
\setcounter{equation}{0}

Several tight non-asymptotic bounds on $P_{\delta}$ (called non-asymptotic equipartition property with respect to conditional entropy) are developed in Appendix \ref{sec:non-aasympt-equip-cond}. The inequalities \eqref{eq-ch4-bms-1}  to \eqref{eq-ch4-bms-6}  can now be established from Theorem \ref{thm-purerandom-bms-1} by applying different upper bounds to $P_{\delta}$ in Theorem \ref{th-nep-cond} in Appendix \ref{sec:non-aasympt-equip-cond}. Towards proving part 1) of this theorem, by \eqref{eq-nep-1} in Theorem \ref{th-nep-cond},
\begin{equation}
  \label{eq-th2-bms-1}
  P_{\delta} \leq \bar{\xi}_H (X|Y, \lambda,n) e^{- n r_{X|Y} (\delta) }
\end{equation}
where $\lambda = r'_{X|Y} (\delta)$. In the meantime, whenever \eqref{eq-ch4-bms-2} holds,
\begin{equation}
  \label{eq-th2-bms-2}
  e^{- n \left( C_{\mathrm{BIMC-L}} -\delta - \mathcal{R} (\mathcal{C}^{(Gal)}_{n,k}) \right)} \leq \lambda \bar{\xi}_H (X|Y, \lambda,n) e^{- n r_{X|Y} (\delta) } .
\end{equation}
Then \eqref{eq-ch4-bms-1} is yielded by plugging \eqref{eq-th2-bms-1} and \eqref{eq-th2-bms-2} into \eqref{eq-th-bms-1} in Theorem \ref{thm-purerandom-bms-1}. The parametric form of $P_e (\mathcal{C}^{(Gal)}_{n,k})$ and $\mathcal{R} (\mathcal{C}^{(Gal)}_{n,k})$ in \eqref{eq-ch4-bms-1} and \eqref{eq-ch4-bms-2} comes from the effort of optimizing $\delta$. Indeed, upon applying \eqref{eq-nep-1} to $P_{\delta}$, the optimal $\delta$ is given by minimizing
\begin{displaymath}
  \bar{\xi}_H (X|Y, \lambda,n) e^{- n r_{X|Y} (\delta) } + e^{-n (C_{\mathrm{BIMC-L}} - \delta - \mathcal{R} (\mathcal{C}^{(Gal)}_{n,k}))}
\end{displaymath}
where the term $\frac{1}{1-2^{-n}}$ is dropped due to its numeric insignificance. Setting  the derivative of above quantity with respect to $\delta$ to zero results in
\begin{equation}
  \label{eq-th2-bms-3}
  \left[ \frac{1}{n} \frac{d  \bar{\xi}_H (X|Y, \lambda,n) }{ d \lambda} \frac{d \lambda}{d \delta} - \lambda \bar{\xi}_H (X|Y, \lambda,n) \right] e^{- n r_{X|Y} (\delta) } + e^{-n (C_{\mathrm{BIMC-L}} - \delta - \mathcal{R} (\mathcal{C}^{(Gal)}_{n,k}))} = 0
\end{equation}
as $\lambda = r'_{X|Y} (\delta)$. To simplify \eqref{eq-th2-bms-3}, $\frac{1}{n} \frac{d  \bar{\xi}_H (X|Y, \lambda,n) }{ d \lambda} \frac{d \lambda}{d \delta} $ is ignored as the magnitude of this term is in general much smaller than $\lambda  \bar{\xi}_H (X|Y, \lambda,n)$ for reasonable values of $n$, and consequently, optimal $\delta$ can be approximated by solving \eqref{eq-ch4-bms-2} or \eqref{eq-th2-bms-2} with equality.

To prove part 2), let $\delta = \frac{c}{ \sqrt{n} }$ and by \eqref{eq-nep-4}, we have
\begin{equation}
  \label{eq-th2-bms-4}
  P_{\frac{c}{\sqrt{n}}} \leq Q \left( \frac{c}{ \sigma_{H} (X|Y) } \right)
    + \frac{1}{\sqrt{n}} \frac{C M_{H} (X|Y)}{\sigma^3_{H} (X|Y)} .
\end{equation}
Meanwhile,
\begin{equation}
  \label{eq-th2-bms-5}
  e^{-n \left( C_{\mathrm{BIMC-L}} -\frac{c}{\sqrt{n}} - \mathcal{R} (\mathcal{C}^{(Gal)}_{n,k}) \right) } \leq \frac{1}{\sqrt{n} \sqrt{2 \pi} \sigma_H (X|Y)} e^{- \frac{c^2}{2 \sigma^2_H (X|Y)} }
\end{equation}
whenever \eqref{eq-ch4-bms-6} is valid. Then \eqref{eq-ch4-bms-5} is proved by combining \eqref{eq-th2-bms-4}, \eqref{eq-th2-bms-5} and \eqref{eq-th-bms-1} in Theorem \ref{thm-purerandom-bms-1}. Similarly, the parametric form of $P_e (\mathcal{C}^{(Gal)}_{n,k})$ and $\mathcal{R} (\mathcal{C}^{(Gal)}_{n,k})$ in \eqref{eq-ch4-bms-5} and \eqref{eq-ch4-bms-6} is yielded by optimizing $c$ to get the tightest bounds as the solution of $c$ to \eqref{eq-ch4-bms-6} or \eqref{eq-th2-bms-5} with equality will minimize
\begin{displaymath}
  Q \left( \frac{c}{ \sigma_{H} (X|Y) } \right) +  \frac{1}{\sqrt{n}} \frac{C M_{H} (X|Y)}{\sigma^3_{H} (X|Y)} + e^{-n \left( C_{\mathrm{BIMC-L}} -\frac{c}{\sqrt{n}} - \mathcal{R} (\mathcal{C}^{(Gal)}_{n,k}) \right) } .
\end{displaymath}

\section{Proof of Corollary \ref{cor-bimc-asym}}
\label{sec:proof-corollary-1}
\setcounter{equation}{0}

When $\epsilon_n = \epsilon$ remains a constant with respect to $n$,
\begin{equation}
  \label{eq-proof-bimc-asym-1}
  \delta_n = O(n^{-0.5})
\end{equation}
and \eqref{eq-cor-bimc-asym-1} can be easily proved by part 2) of Theorem \ref{thm-purerandom-bms}. Now we focus on the case when $\epsilon_n = o(1)$ and $\frac{- \ln \epsilon_n}{n} = o (1)$ as $n \rightarrow +\infty$. In this case, it is easy to verify that $\delta_n = o(1)$ and $\delta_n = \omega (n^{-0.5})$, which further implies that $\frac{1}{n^2 \delta^3_n} = o(\delta_n)$.  Let
\[ \bar{\delta} = \delta_n + d_0 \delta^2_n + \frac{d_1}{n^2 \delta^3_n} \]
for some constants $d_0, d_1 > 0$, and $\bar{\lambda}=r'_{X|Y} (\bar{\delta})$. Now we would like to show that by choosing proper $d_0$ and $d_1$,
\begin{equation}
  \label{eq-proof-bimc-asym-2}
  \left( \frac{1}{1-2^{-n}} + \bar{\lambda} \right) \bar{\xi}_H (X|Y,\bar{\lambda},n) e^{- n r_{X|Y} (\bar{\delta})} \leq \epsilon_n .
\end{equation}
Towards this,
\begin{eqnarray}
  \label{eq-proof-bimc-asym-3}
  \lefteqn{\left( \frac{1}{1-2^{-n}} + \bar{\lambda} \right) \bar{\xi}_H (X|Y,\bar{\lambda},n) e^{- n r_{X|Y} (\bar{\delta})}} \nonumber \\
  &\stackrel{(a)}{\leq}& \left( 1 + \bar{\lambda} + O(2^{-n}) \right) \left( e^{\frac{n \bar{\lambda}^2 \sigma^2_H(X|Y,\bar{\lambda})}{2}} Q(\sqrt{n}  \bar{\lambda} \sigma_H(X|Y,\bar{\lambda})) + \frac{2CM_H(X|Y,\bar{\lambda})}{\sqrt{n} \sigma^3_H(X|Y,\bar{\lambda})} \right) e^{ - n r_{X|Y} (\bar{\delta}) } \nonumber \\
  &\stackrel{(b)}{\leq}& \left( 1 + d_2 \delta_n \right) \left( \frac{1}{\sqrt{2 \pi} \sqrt{n} \bar{\lambda} \sigma_H (X|Y,\bar{\lambda})} + \frac{d_3}{\sqrt{n}} \right) e^{- n \left( \frac{\bar{\delta}^2}{2 \sigma^2_H (X|Y)} - d_4 \bar{\delta}^3 \right) } \nonumber \\
  &\stackrel{(c)}{\leq}& \left( 1 + d_2 \delta_n \right) \left( 1 + d_5 \delta_n \right) \frac{\sigma_H (X|Y)}{\sqrt{2 \pi} \sqrt{n} \delta_n} e^{- n \left( \frac{\delta^2_n}{2 \sigma^2_H (X|Y)} + \left( \frac{d_0}{\sigma^2_H (X|Y)} - d_6 \right) \delta^3_n \right) - \frac{d_1}{\sigma^2_H (X|Y)} \frac{1}{n \delta^2_n} } \nonumber \\
  &\stackrel{(d)}{\leq}& \frac{1}{\sqrt{2 \pi}} \frac{\frac{\sigma_H (X|Y)}{ \sqrt{n} \delta_n}}{1 + \frac{\sigma^2_H (X|Y)}{n \delta^2_n} } e^{- n \left( \frac{\delta^2_n}{2 \sigma^2_H (X|Y)} + \left( \frac{d_0}{\sigma^2_H (X|Y)} - d_6 - d_2 - d_5 \right) \delta^3_n \right) - \left( \frac{d_1}{\sigma^2_H (X|Y)} - \sigma^2_H (X|Y) \right) \frac{1}{n \delta^2_n} } \nonumber \\
  &\stackrel{(e)}{\leq}& \frac{1}{\sqrt{2 \pi}} \frac{\frac{\sigma_H (X|Y)}{ \sqrt{n} \delta_n}}{1 + \frac{\sigma^2_H (X|Y)}{n \delta^2_n} } e^{- \frac{n \delta^2_n}{2 \sigma^2_H (X|Y)}} \nonumber \\
  &\stackrel{(f)}{\leq}& Q \left( \frac{\sqrt{n} \delta_n}{\sigma_H (X|Y)} \right) \stackrel{(g)}{=} \epsilon_n
\end{eqnarray}
where (a) is due to the definition of $\bar{\xi}_H (X|Y,\bar{\lambda},n)$; (b) follows \eqref{eq-nep-a-1} and the fact that
\begin{displaymath}
  \bar{\lambda} = \frac{\bar{\delta}}{\sigma^2_H (X|Y)} + O(\bar{\delta}^2_n) = \frac{\delta_n}{\sigma^2_H (X|Y)} + o(\delta_n)
\end{displaymath}
\begin{displaymath}
  Q(x) \leq \frac{1}{\sqrt{2 \pi} x} e^{- \frac{x^2}{2}}
\end{displaymath}
and $\frac{M_H(X|Y,\lambda)}{ \sigma^3_H(X|Y,\lambda)}$ as a function of $\lambda$ is bounded in a small neighborhood of $\lambda =0$; (c) can be verified by
\begin{eqnarray*}
  \lefteqn{\frac{1}{\sqrt{2 \pi} \sqrt{n} \bar{\lambda} \sigma_H (X|Y,\bar{\lambda})} + \frac{d_3}{\sqrt{n}}} \\
  &=& \frac{\sigma_H (X|Y)}{\sqrt{2 \pi} \sqrt{n} \delta_n} \left( \frac{\delta_n}{\bar{\lambda} \sigma_H (X|Y,\bar{\lambda}) \sigma_H (X|Y)} + O(\delta_n) \right) \\
  &\leq& \frac{\sigma_H (X|Y)}{\sqrt{2 \pi} \sqrt{n} \delta_n} \left( \frac{\delta_n}{\bar{\lambda} \sigma^2_H (X|Y) (1 - O(\bar{\lambda}))} + O(\delta_n) \right) \\
  &=&  \frac{\sigma_H (X|Y)}{\sqrt{2 \pi} \sqrt{n} \delta_n} \left( \frac{\delta_n}{\bar{\lambda} \sigma^2_H (X|Y)} + O(\delta_n) \right) \\
  &\leq& \frac{\sigma_H (X|Y)}{\sqrt{2 \pi} \sqrt{n} \delta_n} \left( \frac{\delta_n}{\bar{\delta} - O(\bar{\delta}^2)} + O(\delta_n) \right) \\
  &\leq& \frac{\sigma_H (X|Y)}{\sqrt{2 \pi} \sqrt{n} \delta_n} \left( \frac{\delta_n}{\delta_n - O(\delta^2_n)} + O(\delta_n) \right) \\
  &=& \frac{\sigma_H (X|Y)}{\sqrt{2 \pi} \sqrt{n} \delta_n} \left( 1 + O(\delta_n) \right)
\end{eqnarray*}
and
\begin{eqnarray*}
   \frac{\bar{\delta}^2}{2 \sigma^2_H (X|Y)} - d_4 \bar{\delta}^3 &=&
    \frac{\left( \delta_n + d_0 \delta^2_n + \frac{d_1}{n^2 \delta^3_n} \right)^2}{2 \sigma^2_H (X|Y)} - O(\delta^3_n) \\
    &\geq& \frac{\delta^2_n + 2d_0 \delta^3_n + \frac{2d_1}{n^2 \delta^2_n} }{2 \sigma^2_H (X|Y)} - d_6 \delta^3_n
\end{eqnarray*}
for some constant $d_6 > 0$; (d) is due to the inequality $e^{x} \geq 1 + x$ and $n \delta^2_n = \omega(1)$; (e) is valid by choosing
\begin{displaymath}
  d_0 = \sigma^2_H (X|Y) (d_2 + d_5 + d_6)
\end{displaymath}
and
\begin{displaymath}
  d_1 = \sigma^4_H (X|Y) ;
\end{displaymath}
(f) follows the inequality
\begin{displaymath}
  \frac{1}{\sqrt{2 \pi}} \frac{x}{1+x^2} e^{-\frac{x^2}{2}} < Q(x)
\end{displaymath}
and (g) is due to the definition of $\delta_n$. Now by part 1) of Theorem \ref{thm-purerandom-bms},
\begin{eqnarray}
  \label{eq-proof-bimc-asym-4}
  \mathcal{R} (\mathcal{C}^{(Gal)}_{n,k}) &\geq& C_{\mathrm{BIMC}} - \bar{\delta} - r_{X|Y} (\bar{\delta}) + \frac{\ln \bar{\lambda} \bar{\xi}_H (X|Y,\bar{\lambda}, n)}{n} \nonumber \\
  &=& C_{\mathrm{BIMC}} - \delta_n - O(\delta^2_n) - O \left( \frac{1}{n^2 \delta^3_n} \right) - \frac{\ln n}{2 n} + \frac{\ln \sqrt{n} \bar{\lambda} \bar{\xi}_H (X|Y,\bar{\lambda}, n)}{n} \nonumber \\
  &=& C_{\mathrm{BIMC}} - \delta_n - O(\delta^2_n) - O \left( \frac{1}{n^2 \delta^3_n} \right) - \frac{\ln n}{2 n} + O(n^{-1})
\end{eqnarray}
where the last step is due to Proposition \ref{cor-nep-asym} in Appendix \ref{sec:non-aasympt-equip-cond}. And the proof of this corollary is completed by observing that
\begin{displaymath}
  O(\delta^2_n) + O \left( \frac{1}{n^2 \delta^3_n} \right) + \frac{\ln n}{2 n} + O(n^{-1}) = o(\delta_n) .
\end{displaymath}

\section{Proof of Theorem \ref{thm-random-dimc-1}}
\label{sec:proof-theorem-3}
\setcounter{equation}{0}

  The proof is along the same way as in the proof of Theorem \ref{thm-purerandom-bms-1}. Let $X^n (q)$ be the transmitted codeword, and  $Y^n$ the output of the DIMC $P$ in response to $X^n (q)$. In parallel with \eqref{eqth2-3}, we have
\begin{eqnarray} \label{eqth3-1}
    P_e (\mathcal{C}_{t, n,k})    & \leq & \Pr \{ X^n (q) \not \in J(Y^n) \} + \Pr \left \{  \exists z^n \neq X^n(q), z^n \in  J(Y^n), z^n  \in \mathcal{C}_{t, n,k}\right\}
\end{eqnarray}
where $J(Y^n)$ is the DIMC jar based on type $t$ as defined in \eqref{eq5-6}. Note that $X^n (q)$ is uniformly distributed over ${\cal T}^n_t$. For any $x^n \in {\cal T}^n_t$ and $y^n \in {\cal Y}^n$, one can verify that
 \begin{eqnarray} \label{eqth3-2}
 \lefteqn{\Pr \left \{ \left.  \exists z^n \neq X^n(q), z^n \in  J(Y^n), z^n  \in \mathcal{C}_{t, n,k} \right | X^n (q) = x^n, Y^n = y^n \right\}} \nonumber \\
    &\stackrel{(a)}{\leq}& |J(y^n)| |\mathcal{T}^n_{t}|^{-1} 2^{k} \nonumber \\
    &\leq&  |J(y^n)| e^{ k \ln 2 - \ln |\mathcal{T}^n_{t}|} \nonumber \\
    &\stackrel{(b)}{\leq}& e^{n \left[ H (t) - I(t;P)  + \delta \right]} e^{ n \left[ \frac{k}{n} \ln 2  \right] - \ln |\mathcal{T}^n_{t}|}   \nonumber \\
    &=& e^{- n \left[ I(t;P) - \delta - \mathcal{R} (\mathcal{C}_{t,n,k}) \right] + n H(t) - \ln |\mathcal{T}^n_{t}|}
    \end{eqnarray}
   where (a)  is due to the fact that all codewords in $\mathcal{C}_{t,n,k}$ are independent, and each is distributed uniformly over ${\cal T}^n_t$, and (b) is verified by
  \begin{eqnarray*}
    |J(y^n)| e^{- n \left( H (t) - I (t;P) + \delta \right)}
    &\leq& \sum_{z^n \in J(y^n)} e^{ -n H (t) + \sum^n_{i=1} \ln \frac{p(y_i|z_i)}{q_{t} (y_i)}} \\
    &=& \sum_{z^n \in J(y^n)} \frac{ e^{ -n H (t)}\prod^n_{i=1} p(y_i|z_i) }{ \prod^n_{i=1} q_{t} (y_i)} \\
    &=& \frac{\sum_{z^n \in J(y^n)}  e^{ -n H (t)}\prod^n_{i=1} p(y_i|z_i) }{ \prod^n_{i=1} \sum_{x \in \mathcal{X}} t(x) p(y_i|x)} \\
    &=& \frac{\sum_{z^n \in J(y^n)}  e^{ -n H (t)}\prod^n_{i=1} p(y_i|z_i) }{ \sum_{x^n \in \mathcal{X}^n} \prod^n_{i=1} t(x_i) p(y_i|x_i)} \\
    &\leq& \frac{\sum_{z^n \in \mathcal{T}^n_{t}}  e^{ -n H (t)}\prod^n_{i=1} p(y_i|z_i) }{ \sum_{x^n \in \mathcal{X}^n} \prod^n_{i=1} t(x_i) p(y_i|x_i)} \\
    &=& \frac{\sum_{z^n \in \mathcal{T}^n_{t}} \prod^n_{i=1} t(z_i) p(y_i|z_i) }{ \sum_{x^n \in \mathcal{X}^n} \prod^n_{i=1} t(x_i) p(y_i|x_i)} \leq 1
  \end{eqnarray*}
  since for any $z^n \in \mathcal{T}^n_{t}$,
  \begin{displaymath}
    \prod^n_{i=1} t(z_i) = e^{-n H (t)}
  \end{displaymath}
  and $\mathcal{T}^n_{t}$ is only a subset of $\mathcal{X}^n$.  Since \eqref{eqth3-2} is valid for any $x^n \in {\cal T}^n_t$ and $y^n \in {\cal Y}^n$, it follows that
  \begin{equation} \label{eqth3-3}
    \Pr \left \{  \exists z^n \neq X^n(q), z^n \in  J(Y^n), z^n  \in \mathcal{C}_{t, n,k} \right\}
    \leq e^{- n \left[ I(t;P) - \delta - \mathcal{R} (\mathcal{C}_{t,n,k}) \right] + n H(t) - \ln |\mathcal{T}^n_{t}|}  \;.
  \end{equation}
  The proof of this theorem is completed by observing that
  \begin{equation}
    \label{eqth3-4}
    \Pr \{ X^n (q) \not \in J(Y^n) \} = P_{t,\delta}
  \end{equation}
  as $X^n (q)$ is drawn from $\mathcal{T}^n_t$.

\section{Comparison with Existing Non-Asymptotic Achievability}
\label{sec:comp-with-exist-achievable-bounds}
\setcounter{equation}{0}

Although there are tremendous achievable bounds \cite{Sason-tutorial,Yury-thesis} (and references therein) on channel coding rate in the prosperous literature of information theory, where various code ensembles and bounding techniques are used, it does not seem that any of our random coding theorems  (Theorems \ref{thm-purerandom-bms-1}, \ref{thm-purerandom-bms}, \ref{thm-random-dimc-1}, and \ref{thm-random-dimc}) could be implied by existing achievability bounds in the literature because of  either the generality of our channel models or the special structure of our random code ensembles in our random coding theorems. For example, Theorems \ref{thm-purerandom-bms-1} and \ref{thm-purerandom-bms} are concerned with Gallager parity check ensemble, wherein codewords are not necessarily pairwise independent, and applicable to  any binary input memoryless channel without any symmetry constraint whatsoever. On the other hand, most achievability bounds on linear block codes are for binary input memoryless channels with symmetry  \cite{Sason-tutorial}. Nonetheless, it is instructive to compare our achievability bounds in Theorems  \ref{thm-purerandom-bms-1}, \ref{thm-purerandom-bms}, \ref{thm-random-dimc-1}, and \ref{thm-random-dimc} with existing bounds in the literature whenever possible. Below we will compare our achievability bounds in Theorems \ref{thm-purerandom-bms-1} and \ref{thm-purerandom-bms} with existing bounds on random linear code ensembles for channels with symmetry, and  our achievability bounds in Theorems  \ref{thm-random-dimc-1}, and \ref{thm-random-dimc} with existing bounds on the existence of codes with a fixed type.

\subsection{Achievability on Random Linear Code Ensembles}
\label{sec:bounds-random-linear}

Random linear code ensembles include Elias generator ensemble and Gallager parity check ensemble. While codewords generated in Elias ensemble are pairwise independent, it is not true for Gallager ensemble. Consequently, non-asymptotic coding theorems on Shannon random code ensemble in the literature, whose proof relies on pairwise independence of codewords, apply only to Elias ensemble, but not to Gallager ensemble. Here we focus on those achievabilities applicable to random linear code ensembles, with the emphasis on Gallager ensemble. Furthermore, as some achievability bounds are only applicable to special channels, we divide our discussion into four parts: 1) bounds for BSCs; 2) bounds for BECs; 3) bounds for BIAGCs; and 4) bounds for MBIOS channels.

\subsubsection{BSC}
\label{sec:ach-bsc}

To make comparison transparent, we rewrite Theorem \ref{thm-purerandom-bms-1}. Let $M=2^k$ be the number of codewords, and $p \in (0,0.5)$ be the crossover probability. By \eqref{eq-th-bms-1} in Theorem~\ref{thm-purerandom-bms-1} and Remark \ref{reth2-1}, it is not hard to verify that
\begin{equation}
  \label{jeq-compare-yury-bsc}
  P_e (\mathcal{C}^{(Gal)}_{n,k})  \leq  \underbrace{\sum_{n \left( p+ \frac{\delta}{\ln \frac{1-p}{p}} \right) < w  \leq n}
  \left(
  \begin{array}{c}
    n \\
    w
  \end{array}
  \right) p^w (1-p)^{n-w}}_{\Pr \left\{ X^n \notin J(Y^n) \right\} }  +
  \sum_{0 \leq w \leq n \left( p+ \frac{\delta}{\ln \frac{1-p}{p}} \right)}
  \left(
  \begin{array}{c}
    n \\
    w
  \end{array}
  \right) 2^{-n} M .
\end{equation}
Further optimizing $\delta$ implies that
\begin{equation}
  \label{jeq-compare-yury-bsc-op}
  P_e (\mathcal{C}^{(Gal)}_{n,k}) \leq \sum^n_{w =0}
  \left(
  \begin{array}{c}
    n \\
    w
  \end{array}
  \right) \min \left\{ p^w (1-p)^{n-w}, 2^{-n} M \right\}
\end{equation}
and \eqref{jeq-compare-yury-bsc-op} is essentially the same (except for a minor difference\footnote{Replacing $M$ in \eqref{jeq-compare-yury-bsc-op} by $(M-1)/2$ yields exactly the Dependence Testing Bound \cite[Theorem 34]{Yury-Poor-Verdu-2010}.}) as the Dependence Testing Bound recently established in \cite[Theorem 34]{Yury-Poor-Verdu-2010} for Shannon random code ensemble and Elias ensemble over the BSC.

As discussed in the introduction section, Poltyrev derived an achievability bound for any deterministic code in terms of its Hamming weight profile $\{ N(l) \}^n_{l=1}$ on BSCs, and by replacing $N(l)$ with $2^{-(n-k)} {n \choose l}$, the resulting bound holds for Gallager ensemble $\mathcal{C}^{(Gal)}_{n,k}$, as well as Elias ensemble. In addition, it was shown that Random Coding Union Bound \cite[Theorem 33]{Yury-Poor-Verdu-2010} derived for Shannon random code ensemble and Elias ensemble is the same as Poltyrev's bound.

\begin{figure}[h!]
  \centering
  \subfloat[$P_e = 10^{-3}$]{\includegraphics[scale=0.4]{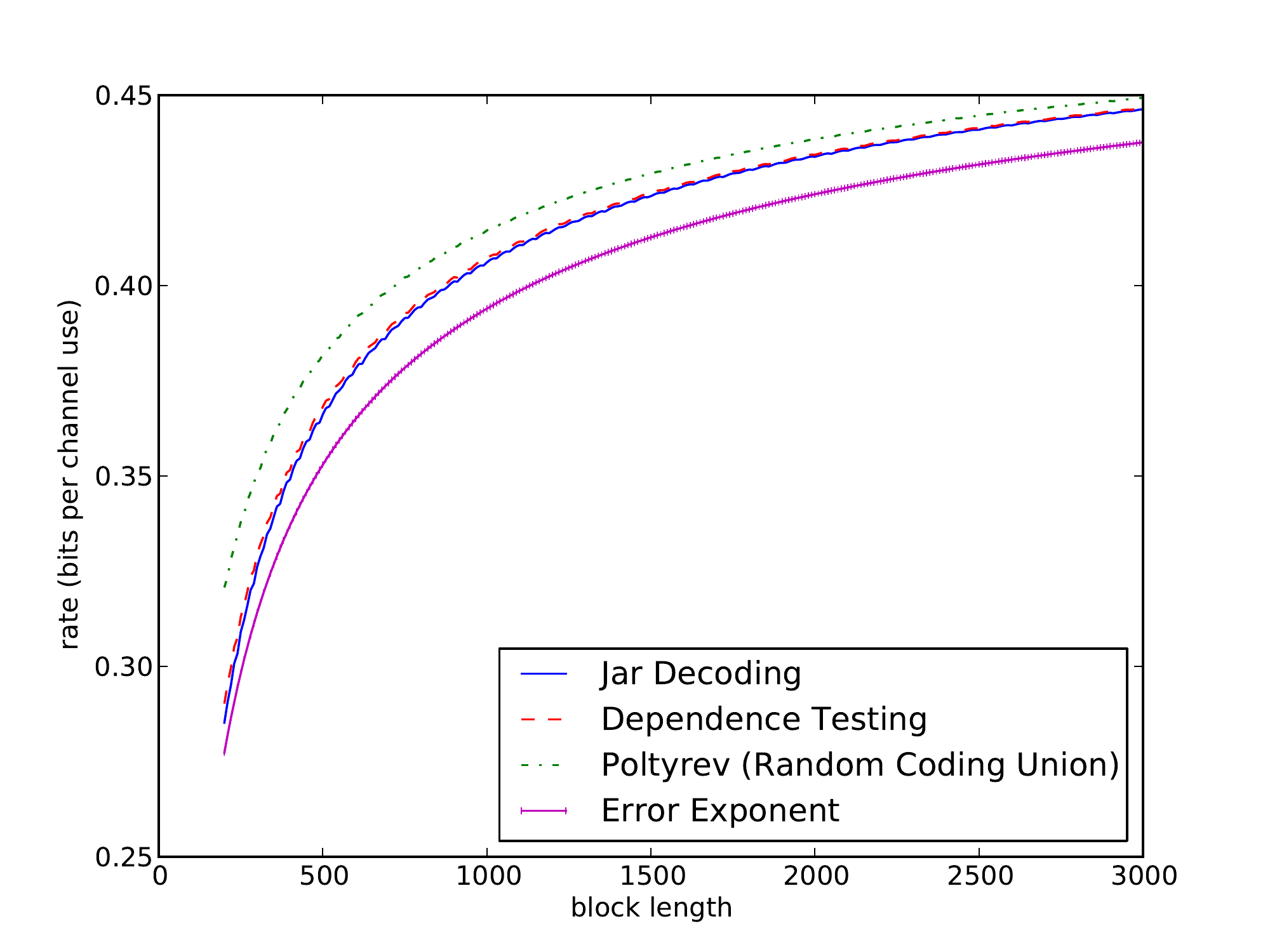}}
  \subfloat[$P_e = 10^{-6}$]{\includegraphics[scale=0.4]{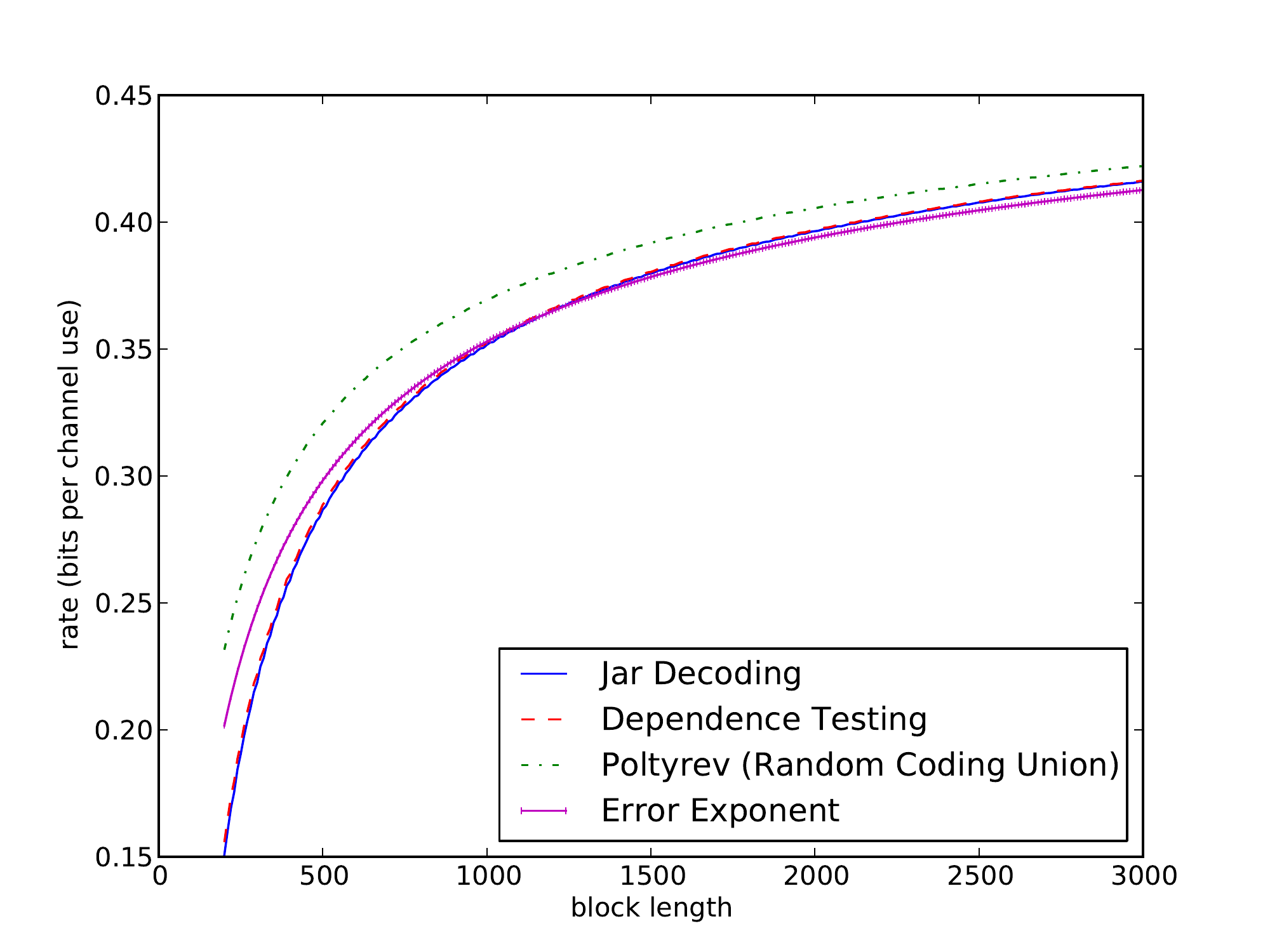}}
  \caption{Comparison of Achievability for BSC with cross-over probability $p=0.11$}
  \label{fig:ach-bsc}
\end{figure}
Figure \ref{fig:ach-bsc} shows the numeric comparison (with block length range $[200,3000]$ and fixed word error probability $10^{-3}$ and $10^{-6}$) among Theorem \ref{thm-purerandom-bms-1}, Poltyrev's Bound \cite[Lemma 1]{Poltyrev} (Random Coding Union Bound \cite[Theorem 33]{Yury-Poor-Verdu-2010}) and Error Exponent Bound on a BSC with cross-over probability $p=0.11$, where Dependence Testing Bound \cite[Theorem 34]{Yury-Poor-Verdu-2010} is also included for a benchmark. As can be seen, the numeric result confirms that Theorem \ref{thm-purerandom-bms-1} is essentially the same as Dependence Testing Bound and further shows that Poltyrev's Bound (Random Coding Union Bound) is better than Dependence Testing Bound and Theorem \ref{thm-purerandom-bms-1} by a small margin, while Dependence Testing Bound and Theorem \ref{thm-purerandom-bms-1} outperform Error Exponent Bound when word error probability is relatively large with respect to block length, which is consistent with the observation in \cite{Yury-Poor-Verdu-2010}.

\subsubsection{BEC}
\label{sec:ach-bec}

Now let us focus on a BEC. In this case, Theorem \ref{thm-purerandom-bms-1} can be further improved as follows. Let $M=2^k$ be the number of codewords and $p$ be the erasure probability. It is then easy to verify that
\begin{displaymath}
  H (X|Y) = p \ln 2
\end{displaymath}
and in this case, the BIMC-L jar reduces to
\begin{displaymath}
  J(y^n) =
  \left\{
    \begin{array}{ll}
      \left\{x^n: x_i = y_i \mbox{ if $y_i \neq e$} \right\} &
      \mbox{if $|\left\{ i : y_i = e \right\}| \leq n \left( p +
          \frac{\delta}{\ln 2} \right)$} \\
      \mbox{empty} & \mbox{otherwise}
    \end{array}
  \right. .
\end{displaymath}
Following the argument in the proof of Theorem \ref{thm-purerandom-bms-1}, we have
\begin{eqnarray}
  \label{jeq-compare-yury-bec}
    P_e (\mathcal{C}^{(Gal)}_{n,k}) & \leq & \underbrace{\sum_{n(p+\frac{\delta}{\ln 2}) < t \leq n}
  \left(
  \begin{array}{c}
    n \\
    t
  \end{array}
  \right) p^t (1-p)^{n-t}}_{\Pr \left\{ X^n (q) \notin J(Y^n) \right\} }  \nonumber \\
  & & \mbox{ } +
  \Pr \left\{ \exists z^n \neq X^n(q), z^n \in
    J(Y^n), z^n \in \mathcal{C}^{(Gal)}_{n,k} \right\}  \nonumber \\
  & \leq &  \sum_{n(p+\frac{\delta}{\ln 2}) <  t \leq n}
  \left(
  \begin{array}{c}
    n \\
    t
  \end{array}
  \right) p^t (1-p)^{n-t}  \nonumber \\
  & & \mbox{ } +
  \sum_{1 \leq t \leq n(p+\frac{\delta}{\ln 2})}
  \left(
  \begin{array}{c}
    n \\
    t
  \end{array}
  \right) p^{t} (1-p)^{n-t} 2^{t} 2^{-n} M
\end{eqnarray}
and optimizing $\delta$ yields
\begin{eqnarray}
  \label{jeq-compare-yury-bec-op}
  P_e (\mathcal{C}^{(Gal)}_{n,k}) &\leq&  \sum^n_{t=1}
  \left(
  \begin{array}{c}
    n \\
    t
  \end{array}
  \right) p^t (1-p)^{n-t} \min \left\{ 1, 2^{-(n-t)}M \right\}
  \nonumber \\
  &=& \sum^n_{t=1}
  \left(
  \begin{array}{c}
    n \\
    t
  \end{array}
  \right) p^t (1-p)^{n-t} 2^{-\left[ n - t - \log_2 M \right]^+}
\end{eqnarray}
which is again essentially the same (except for a minor difference\footnote{Replacing $M$ by $(M-1)/2$, and then starting the summation from $t=0$ instead of $t=1$ in \eqref{jeq-compare-yury-bec-op} yield exactly the Dependence Testing Bound \cite[Theorem 37]{Yury-Poor-Verdu-2010}.}) as the Dependence Testing Bound \cite[Theorem 37]{Yury-Poor-Verdu-2010} for Shannon random code ensemble and Elias generator ensemble.  Note that $\frac{1}{1-2^{-n}}$ in Theorem \ref{thm-purerandom-bms-1} is dropped here according to Remark \ref{reth2-1}.

\begin{figure}[h!]
  \centering
  \subfloat[$P_e = 10^{-3}$]{\includegraphics[scale=0.3]{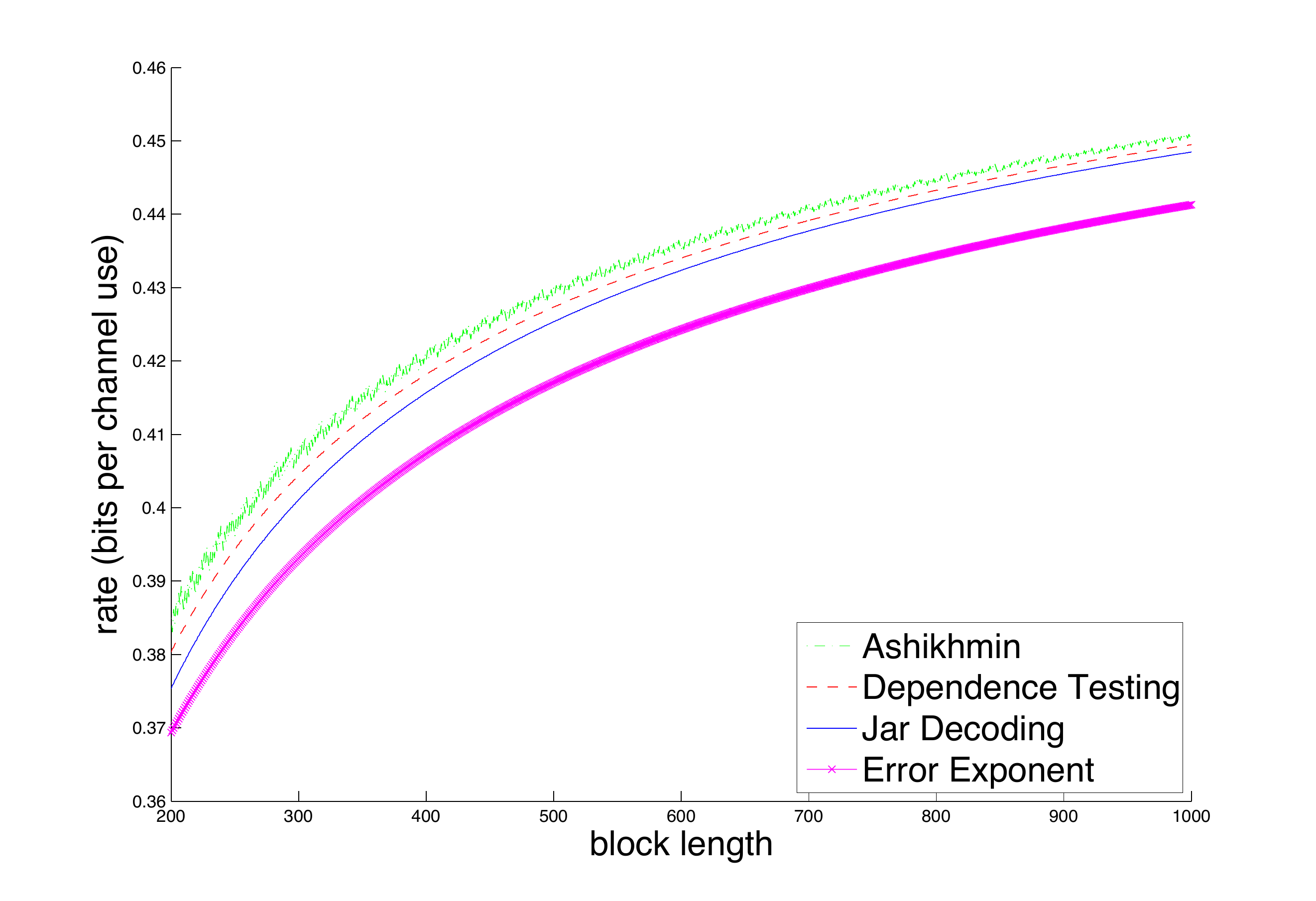}}
  \subfloat[$P_e = 10^{-6}$]{\includegraphics[scale=0.3]{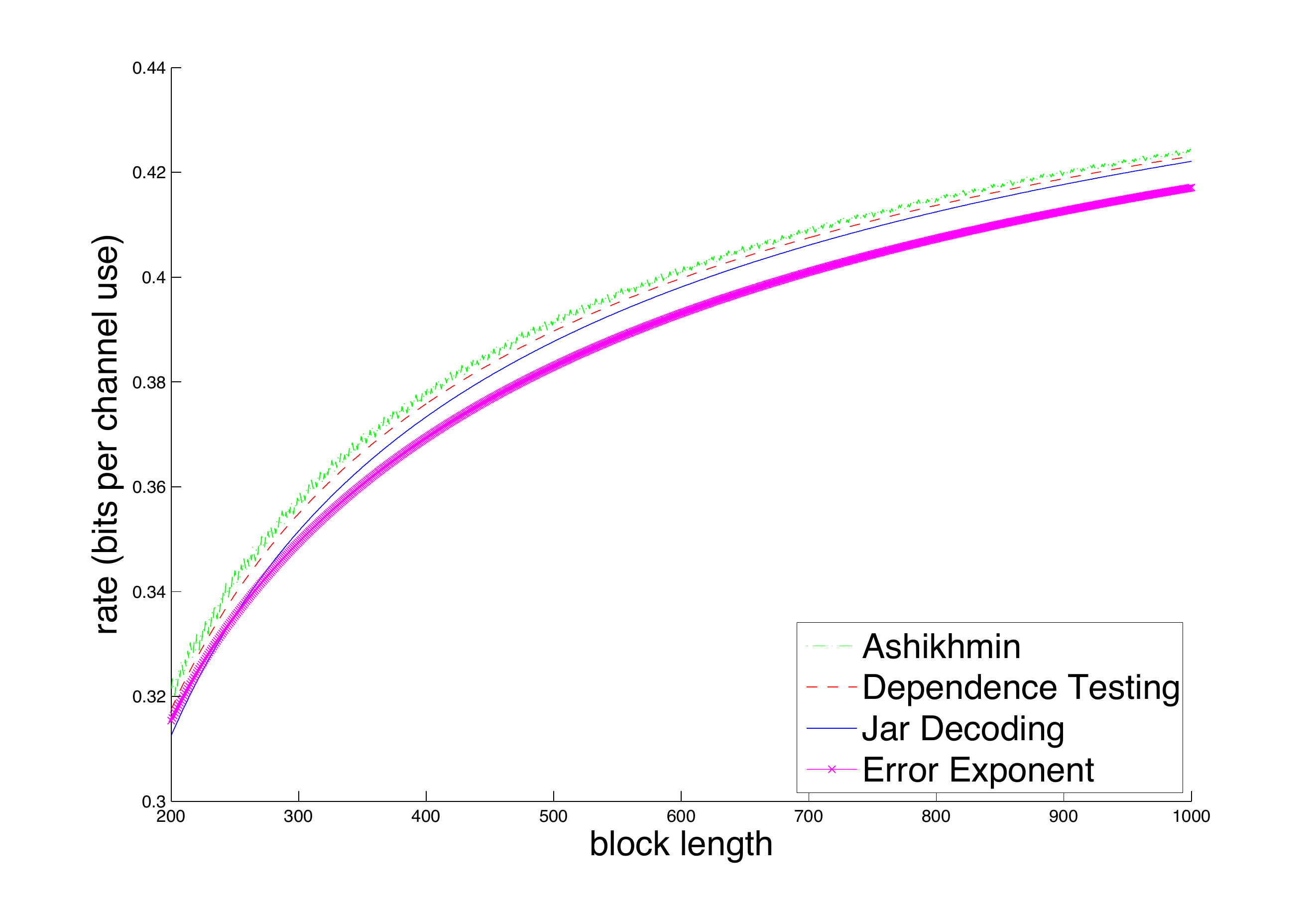}}
  \caption{Comparison of Achievability for BEC with erasure probability $p=0.5$}
  \label{fig:ach-bec}
\end{figure}

For BECs, Ashikmin derived an expression for word error probability of full rank Elias ensemble (i.e. the generator matrix is equiprobably selected among all full rank matrices), included as Theorem 6 in \cite{Yury-Poor-Verdu-2010}. 
Figure \ref{fig:ach-bec} shows the numeric comparison among (\ref{jeq-compare-yury-bec-op}), Ashikmin's Bound, Error Exponent Bound, and Dependence Testing Bound \cite[Theorem 37]{Yury-Poor-Verdu-2010}. Once again, our achievability is very close to Dependence Testing Bound, outperforms Error Exponent Bound, and is worse than Ashikmin's Bound (the best achievability under ML decoding known so far) by a small margin.

\subsubsection{BIAGC}
\label{sec:ach-biagc}

Since in this case, there is no feasible way to calculate $P_{\delta}$, we apply part 1) of Theorem \ref{thm-purerandom-bms}, where $\frac{1}{1-2^{-n}}$ in \eqref{eq-ch4-bms-1} is replaced by $1$ due to Remark \ref{reth2-1}

\begin{figure}[h!]
  \centering
  \includegraphics[scale=0.3]{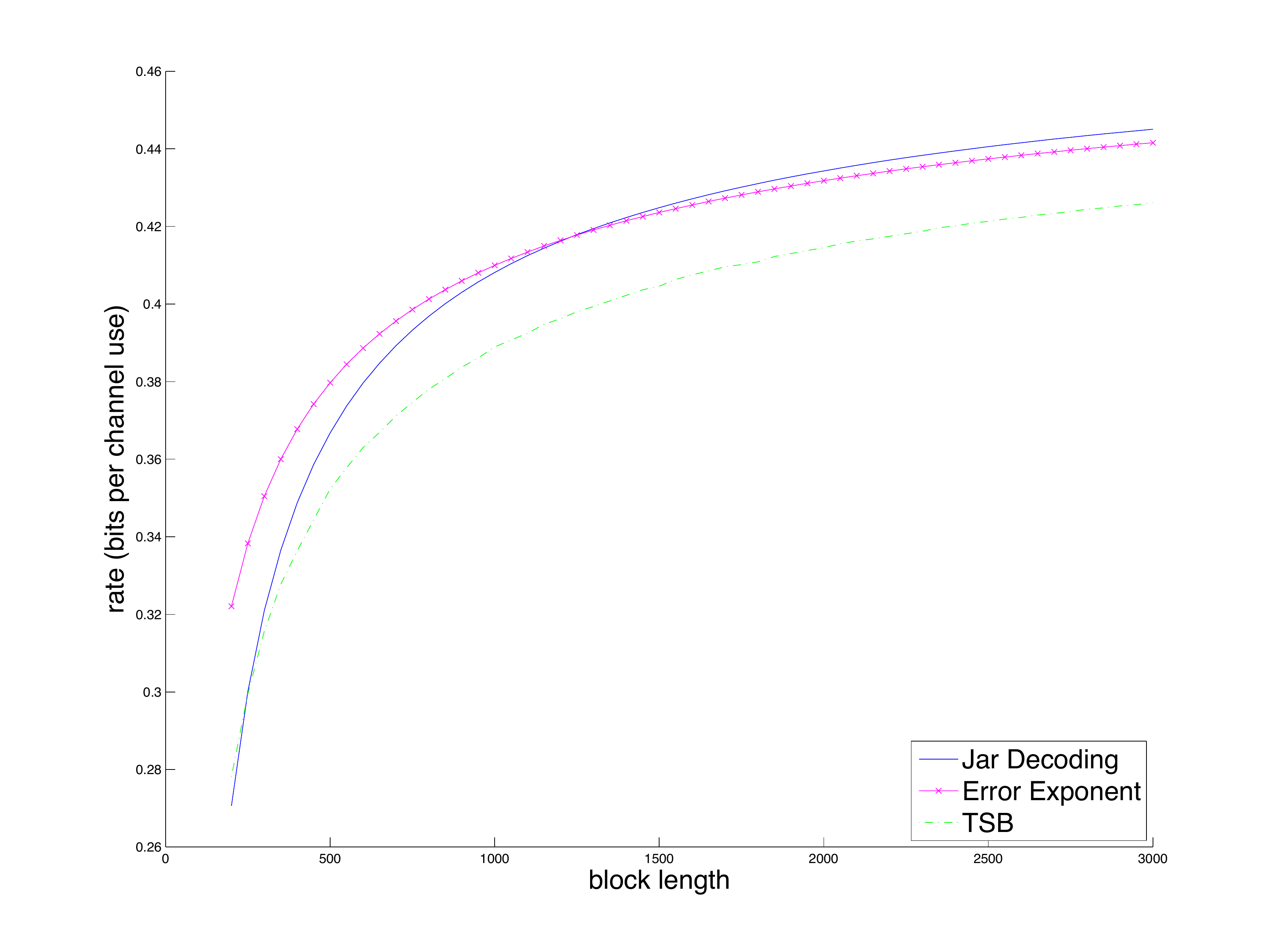}
  \caption{Comparison of Achievability for BIAGC with snr $0$dB and word error probability $P_e = 10^{-2}$}
  \label{fig:ach-biagc}
\end{figure}

There is a rich literature about error probability bounds of linear codes for BIAGCs. One of the tightest bounds in this research area is TSB, proved by Poltyrev in \cite{Poltyrev}. TSB was then improved by Yousefi and Khandani in \cite{yousefi-1}, and Mehrabian and Yousefi in \cite{yousefi-2}. It is unclear, however, whether those two improved bounds can be efficiently evaluated for Gallager parity check ensemble.  Although TSB is one of the tightest bounds for any deterministic code in terms of its Hamming weight profile, it fails to reproduce the Gallager error exponent (\cite{Sason-tutorial} and references therein ) for Gallager parity check ensemble.  Figure \ref{fig:ach-biagc} shows numerical comparison among part 1) of Theorem \ref{thm-purerandom-bms} (~\eqref{eq-ch4-bms-1} and \eqref{eq-ch4-bms-2} ), TSB,  and Error Exponent Bound, where the signal-to-noise ratio (snr) is $0$dB and the word error probability is kept to be $10^{-2}$. As can be seen, TSB is worse than Error Exponent Bound, while our achievability is better than Error Exponent Bound in certain block length region. To the best of our knowledge, this is the first numeric demonstration that Error Exponent Bound can be beaten in the non-asymptotic regime for BIAGCs as well.

\subsubsection{General MBIOS Channels}
\label{sec:ach-mibos-channel}

The only existing achievability bound in the literature applicable to this general case is
Error Exponent Bound for Gallager ensemble, as well as Elias ensemble. The symmetry property of MBIOS channels
is essential to the proof of Error Exponent Bound for Gallager ensembles. As demonstrated already, our achievability bounds in Theorems \ref{thm-purerandom-bms-1} and \ref{thm-purerandom-bms}, applicable to any BIMC, can be tighter than Error Exponent Bound in the non-asymptotic regime.

\subsubsection{Summary}
Applicability (to ensembles and channels) and computational complexity of jar decoding achievability and existing achievability bounds for random linear code ensembles in the literature are summarized in Table \ref{tab:ach-bimc}, where by unknown, we means that at this point we are not aware of any method which can be used to effectively compute the corresponding bound. Among all the listed results, Theorem \ref{thm-purerandom-bms} is the only achievability that can be applied to general BIMCs and efficiently evaluated. Focusing on Gallager ensemble, existing achievability bounds only deal with MBIOS channels, which are a strict subset of BIMCs. For some special MBIOS channels, e.g. BSCs and BECs, there are bounds proved under ML decoding, which are better than our achievability in \eqref{jeq-compare-yury-bsc-op} and \eqref{jeq-compare-yury-bec-op} by a small margin in the non-asymptotic regime. For general MBIOS channels, however, to the best of our knowledge, Error Exponent Bound was the best computable achievability result in the literature before this paper. And numerical calculation shows that the achievability bound in Theorem \ref{thm-purerandom-bms} can be tighter than Error Exponent Bound in the non-asymptotic regime.

\begin{table}[h!]
  \centering
  \begin{tabular}{|c|c|c|c|c|}
    \hline
    \multicolumn{2}{|c|}{\multirow{2}{*}{Achievability Bounds}} & \multicolumn{2}{c|}{Applicability} & Computational \\
    \cline{3-4}
     \multicolumn{2}{|c|}{} & Linear Code Ensembles & BIMC & Complexity \\
    \hline
    \multirow{3}{*}{Jar Decoding} & \eqref{jeq-compare-yury-bsc-op} & \multirow{3}{*}{$\surd$ Elias $\surd$ Gallager} & BSC & $O(n)$ \\
    \cline{2-2} \cline{4-5}
    & \eqref{jeq-compare-yury-bec-op} & & BEC & $O(n)$ \\
    \cline{2-2} \cline{4-5}
    & Theorem \ref{thm-purerandom-bms} & & General & $O(1)$ \\ \hline
    Poltyrev & \cite[Lemma 1]{Poltyrev} & $\surd$ Elias $\surd$ Gallager & BSC & $O(n)$ \\
    \hline
    Ashikmin & \cite[Theorem 6]{Yury-Poor-Verdu-2010} & $\surd$ Elias (full rank) $\times$ Gallager & BEC & $O(n^2)$ \\
    \hline
    TSB & \cite[Lemma 4]{Poltyrev}  & $\surd$ Elias $\surd$ Gallager & BIAGC & $O(1)$ \\
    \hline
    Error Exponent & \cite{shu-feder1999} & $\surd$ Elias $\surd$ Gallager & MBIOS & $O(1)$ \\
    \hline
    \multirow{2}{*}{Random Coding Union} & \cite[Theorem 33]{Yury-Poor-Verdu-2010} & \multirow{2}{*}{$\surd$ Elias $\times$ Gallager} & BSC & $O(n)$\\
    \cline{2-2} \cline{4-5}
    & \cite[Theorem 16]{Yury-Poor-Verdu-2010} & & General & Unknown \\
    \hline
    \multirow{3}{*}{Dependence Testing} & \cite[Theorem 34]{Yury-Poor-Verdu-2010} & \multirow{3}{*}{$\surd$ Elias $\times$ Gallager} & BSC & $O(n)$ \\
    \cline{2-2} \cline{4-5}
    & \cite[Theorem 37]{Yury-Poor-Verdu-2010} & & BEC & $O(n)$ \\
    \cline{2-2} \cline{4-5}
    & \cite[Theorem 17]{Yury-Poor-Verdu-2010} & & General & Unknown \\
    \hline
  \end{tabular}
  \caption{Achievability bounds of Random Linear Codes for BIMCs}
  \label{tab:ach-bimc}
\end{table}

\subsection{Achievability on Shannon Random Code Ensemble With a Fixed Codeword Type}
\label{sec:bounds-random-code}

Technically speaking, when channel input is discrete, achievability results for Shannon random code ensemble also apply to the code ensemble with a fixed codeword type $t$, by restricting the input distribution in $\mathcal{T}^n_t$. In this case, however, neither the input nor output distribution has the product form. Consequently, the evaluation of those achievability bounds becomes much more challenging. In contrast, our achievability in Theorem \ref{thm-random-dimc-1} can be always easily computed for DIMCs with discrete output, while Theorem \ref{thm-random-dimc} can be used when the channel output is continuous. Therefore, in this subsection, we focus on those achievability bounds on random code ensemble with a fixed codeword type, which allow efficient evaluation.

Reviewing results in the literature, a connection between Theorem \ref{thm-random-dimc-1} and $\kappa \beta$ bound \cite[Theorem 25]{Yury-Poor-Verdu-2010} is found. Towards showing this connection, the following definitions are needed. Let $q_1 (w^n)$ and $q_2 (w^n)$,  $w^n \in \mathcal{W}^n$,  be two distributions on a sample space $\mathcal{W}^n$, and $p_{Z|W^n} (z|w^n)$  be a distribution over $z \in \{0,1\}$ given any $w^n \in \mathcal{W}^n$. Define for $\alpha \in [0,1]$
\begin{equation}
  \label{eq-beta-defn-1}
  \beta_{\alpha} (q_1, q_2) \defeq \min_{p_{Z|W^n}: \int q_1 (w^n) p_{Z|W^n} (1|w^n) dw^n \geq \alpha} \int q_2 (w^n) p_{1|W^n} (1|w^n) dw^n.
\end{equation}
In hypothesis testing, the conditional distribution $p^*_{Z|W^n}$ achieving the above optimization can be interpreted as an optimal randomized test  between $q_1$ (null) and $q_2$ (alternative). Now given any distribution $q_{Y^n} (y^n)$  over $y^n \in \mathcal{Y}^n$ and conditional distribution  $p_{Y^n|X^n=x^n} (y^n) \defeq p_{Y^n|X^n} (y^n|x^n)$ over $y^n \in \mathcal{Y}^n$ given any  $x^n \in \mathcal{X}^n$,  further define for $\alpha \in [0,1]$
\begin{equation}
  \label{eq-beta-defn-2}
  \beta_{\alpha} (x^n,q_{Y^n}) \defeq \beta_{\alpha} (p_{Y^n|X^n=x^n}, q_{Y^n}).
\end{equation}
 In addition, for $\mathcal{F} \subseteq \mathcal{X}^n$ and $\tau \in [0,1]$, define
\begin{equation}
  \label{eq-kappa-defn}
  \kappa_{\tau} (\mathcal{F}, q_{Y^n}) = \inf_{p_{Z|Y^n} : \inf\limits_{x^n \in \mathcal{F}} \int p_{Y^n|X^n} (y^n|x^n) p(1|y^n) dy^n \geq \tau} \int q_{Y^n} (y^n) p_{Z|Y^n} (1|y^n) dy^n.
\end{equation}
Then the following result is proved in \cite{Yury-Poor-Verdu-2010}.
\begin{result}[$\kappa \beta$  Bound {\cite[Theorem 25]{Yury-Poor-Verdu-2010}}]
  \label{result-kappabeta}
  Given any channel $\{ p_{Y^n|X^n} (y^n|x^n) : x^n \in \mathcal{X}^n, y \in \mathcal{Y}^n \}$ and $\mathcal{F} \subseteq \mathcal{X}^n$, there exists  a channel code $\mathcal{C}_n$ with  $M$ codewords, all of which are from $\mathcal{F}$, satisfying
  \begin{equation}
    \label{eq-kappabeta-1}
    M \geq \sup_{0 < \tau < P_e (\mathcal{C}_n)} \sup_{q_{Y^n}}
    \frac{\kappa_{\tau} (\mathcal{F}, q_{Y^n})}{\sup\limits_{x^n \in \mathcal{F}} \beta_{1- P_e (\mathcal{C}_n)+\tau} (x^n,q_{Y^n})} .
  \end{equation}
\end{result}

 In general, $\beta$ and $\kappa$ defined above are difficult to evaluate.  Upper and lower bounds on $\beta$ and $\kappa$ are provided in \cite[Equations (103), (104), (106), (121) and (122)]{Yury-Poor-Verdu-2010}, and included here for easy reference:
\begin{equation}
  \label{eq-beta-ub}
  \beta_{\alpha} (q_1,q_2) \leq \frac{1}{\sup\limits_{\gamma : \Pr \left\{ \frac{q_1 (W^n)}{q_2 (W^n)} \geq \gamma \right\} \geq \alpha } \gamma }
\end{equation}
where $W^n$ follows the distribution $q_1$,
\begin{equation}
  \label{eq-beta-lb}
  \beta_{\alpha} (x^n,q_{Y^n}) \geq \sup_{\gamma > 0} \frac{1}{\gamma} \left( \alpha - \Pr \left\{ \frac{p_{Y^n|X^n} (Y^n | x^n)}{ q_{Y^n} (Y^n)}  \geq \gamma \right\} \right)
\end{equation}
where $Y^n$ follows the distribution $p_{Y^n|X^n=x^n}$ given $x^n$, and
\begin{equation}
  \label{eq-kappa-b}
  \tau \int_{x^n \in \mathcal{F}} p_{X^n} (x^n) dx^n \leq \kappa_{\tau} (\mathcal{F}, q_{Y^n}) \leq \tau
\end{equation}
when $ q_{Y^n}$ satisfies
\begin{displaymath}
  q_{Y^n} (y^n) = \int p_{X^n} (x^n) p_{Y^n|X^n} (y^n|x^n) dx^n.
\end{displaymath}
Now let us compare Theorem \ref{thm-random-dimc-1} and Result \ref{result-kappabeta}. Strictly speaking, Result \ref{result-kappabeta} is not applicable to Shannon random code ensemble with a fixed codeword type, as its proof constructs a channel code in a greedy, deterministic way.  Nevertheless, both Theorem \ref{thm-random-dimc-1} and Result \ref{result-kappabeta} imply the existence of channel codes with certain property and performance. Specifically, give a type $t$, let $\mathcal{F} = \mathcal{T}^n_t$ and $q_{Y^n} (y^n) = q_t (y^n) = \prod^n_{i=1} q_t (y_i)$. It is then easy to verify that $\beta_{\alpha} (x^n, q_t)$ is a constant (denoted by $\beta_{\alpha} (q_t)$ ) depending on  $x^n \in \mathcal{F}$ only through its type $t$. Consequently, the bound (\ref{eq-kappabeta-1}) reduces to
\begin{equation}
  \label{eq-kappabeta-2}
  M \geq \sup_{0 < \tau < P_e (\mathcal{C}_{t,n,k})} \frac{\kappa_{\tau} (\mathcal{T}^n_t, q_t)}{\beta_{1-P_e (\mathcal{C}_{t,n,k})+\tau} (q_t)} .
\end{equation}
 From (\ref{eq-kappa-b}) and (\ref{eq-beta-ub}), it follows that  
\begin{equation}
  \label{eq-kappabeta-3}
  \kappa_{\tau} (\mathcal{T}^n_t, q_t) \geq \tau e^{- n H(t)} |\mathcal{T}^n_t|
\end{equation}
and $\forall x^n \in \mathcal{T}^n_t$,
\begin{eqnarray}
  \label{eq-kappabeta-4}
  \frac{1}{\beta_{1-P_e (\mathcal{C}_{t,n,k})+\tau} (x^n, q_t)} &\geq& \sup \left\{ \gamma : \Pr \left\{ \frac{p(Y^n|x^n)}{q_t (Y^n)} \geq \gamma \right\} \geq 1 - P_e (\mathcal{C}_{t,n,k}) + \tau \right\} \nonumber \\
  &=& \sup \left\{ e^{\gamma} : \Pr \left\{ \ln \frac{p(Y^n|x^n)}{q_t (Y^n)} < \gamma \right\} \leq  P_e (\mathcal{C}_{t,n,k}) - \tau \right\} \nonumber \\
  &=& \sup_{\delta : P_{t,\delta} \leq P_e (\mathcal{C}_{t,n,k}) - \tau} e^{I(t;P) - \delta}
\end{eqnarray}
where $Y^n$ is the channel response to $x^n$. Now plugging (\ref{eq-kappabeta-3}) and (\ref{eq-kappabeta-4}) into (\ref{eq-kappabeta-2}), taking logarithm and then dividing $n$ on both sides, we get
\begin{eqnarray}
  \label{eq-kappabeta-5}
  \mathcal{R} (C_{t,n,k}) &\geq& \sup_{0 < \tau < P_e (\mathcal{C}_{t,n,k})} \sup_{\delta : P_{t,\delta} \leq P_e (\mathcal{C}_{t,n,k}) - \tau} I(t;P) - \delta + \frac{\ln \tau + \ln e^{- n H(t)} |\mathcal{T}^n_t|}{n} \nonumber \\
  &=& \sup_{\delta : P_{t,\delta} < P_e (\mathcal{C}_{t,n,k})} \sup_{0 < \tau \leq P_e (\mathcal{C}_{t,n,k}) - P_{t,\delta}} I(t;P) - \delta + \frac{\ln \tau + \ln e^{- n H(t)} |\mathcal{T}^n_t|}{n} \nonumber \\
  &=&  \sup_{\delta : P_{t,\delta} < P_e (\mathcal{C}_{t,n,k})} I(t;P) - \delta + \frac{\ln \left( P_e (\mathcal{C}_{t,n,k}) - P_{t,\delta} \right) + \ln e^{- n H(t)} |\mathcal{T}^n_t|}{n}
\end{eqnarray}
which is equivalent to \eqref{eq-th-random-dimc-1} in Theorem \ref{thm-random-dimc-1}. Consequently, both Result \ref{result-kappabeta} and Theorem \ref{thm-random-dimc-1} imply the existence of a channel code with a fixed codeword $t$ achieving the trade-off between the rate and the word error probability in \eqref{eq-th-random-dimc-1}. And both of the results go beyond this existence in their own ways. Result \ref{result-kappabeta} holds for maximal error probability, and the achievability (\ref{eq-kappabeta-1}) might be tighter than \eqref{eq-th-random-dimc-1} in general, although the evaluation of $\beta$ and $\kappa$ is quite challenging. Theorem \ref{thm-random-dimc-1}, on the other hand, shows that the average coding performance (the rate and the word error probability) of random code ensemble with a fixed codeword type can achieve \eqref{eq-th-random-dimc-1}, which implies the existence result, but not vice versa.

\begin{figure}[h]
  \centering
  \begin{tikzpicture}[rounded corners,ultra thick]
    \path (0,0) node(X0) {$0$} (5.5,0) node(Y0) {$0$}
             (0,1) node(X)   {$X$} (5.5,1) node(Y)   {$Y$}
             (0,2) node(X1) {$1$} (5.5,2) node(Y1) {$1$};

    \draw[->] (X0) -- (Y0) node[above,pos=0.5] {$1-p$};
    \draw[->] (X0) -- (Y1) node[above,pos=0.5] {$p$};
    \draw[->] (X1) -- (Y1) node[above,pos=0.5] {$1$};
  \end{tikzpicture}
  \caption{Z Channel}
  \label{fig:zch}
\end{figure}
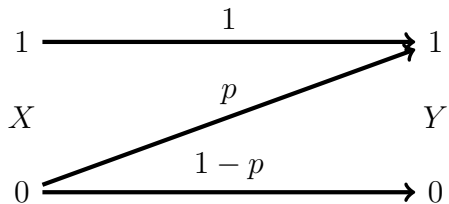

Next, we move on to the error exponent result, proved by Fano in \cite{fano1961} on any discrete (input and output) memoryless channel (DMC). Particularly, Fano showed that given a DMC and a type $t$, the error exponent achieved by Shannon random code ensemble with a fixed codeword type $t$ is larger than that achieved by Shannon random code ensemble with input distribution $t$ in general. Towards numeric comparison between Fano's result and Theorem \ref{thm-random-dimc-1}, we consider a special DIMC with discrete output, Z channel, shown in Figure \ref{fig:zch}.
As can be seen, Z channel and BEC share some common properties. Consequently, the achievability in Theorem \ref{thm-random-dimc-1} can be further improved by providing a better bound on the size of jar $|J(y^n)| $ given a channel output $y^n$. Given a type $t$, the improved achievability is shown below
\begin{equation}
  \label{eq-nep-z-5}
  P_e (\mathcal{C}_{t,n,k}) \leq \sum^{m}_{i=0}
    \left(
      \begin{array}{c}
        m \\
        i
      \end{array}
    \right) (1-p)^{m-i} p^i \min \left\{ 1, (M-1)
              \frac{ \left(
                \begin{array}{c}
                  n-m+i \\
                  i
                \end{array}
                     \right)}
                     { \left(
                \begin{array}{c}
                  n \\
                  m
                \end{array}
                     \right)}
    \right\}
\end{equation}
where $M=2^{n \mathcal{R}(\mathcal{C}_{t,n,k})}$ and $m=t(0) n$. Then (\ref{eq-nep-z-5}) (Jar Decoding) is numerically compared with Fano's result on Z channel with different channel parameters $p$ and input types $t$, where Gallager's Error Exponent Bound on Shannon random code ensemble with input distributions corresponding to $t$ serves as a benchmark.

\begin{figure}[h!]
  \centering
  \subfloat[$t = ( 0.5, 0.5 )$]{\includegraphics[scale=0.3]{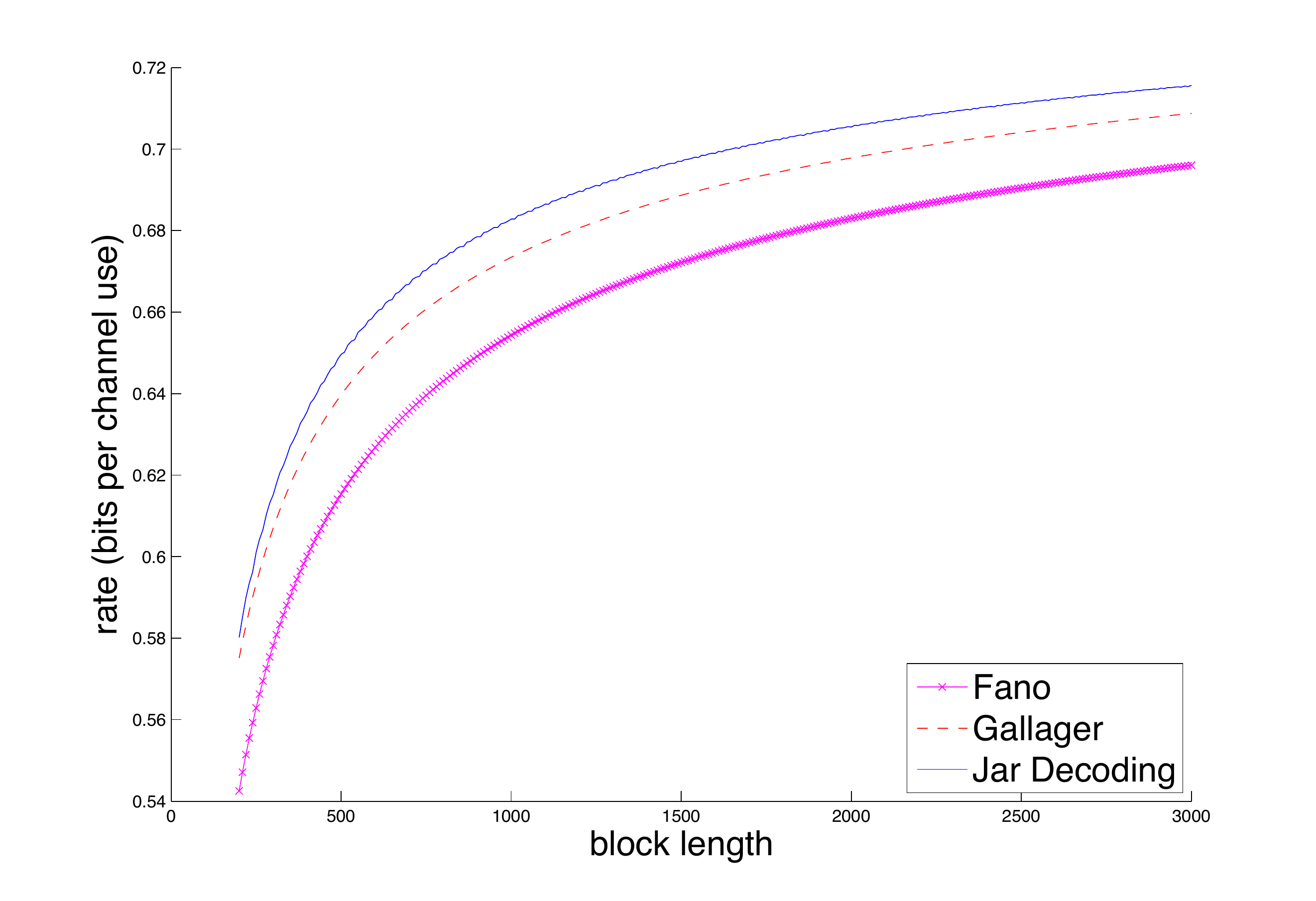}}
  \subfloat[$t = ( 0.1, 0.9 )$]{\includegraphics[scale=0.3]{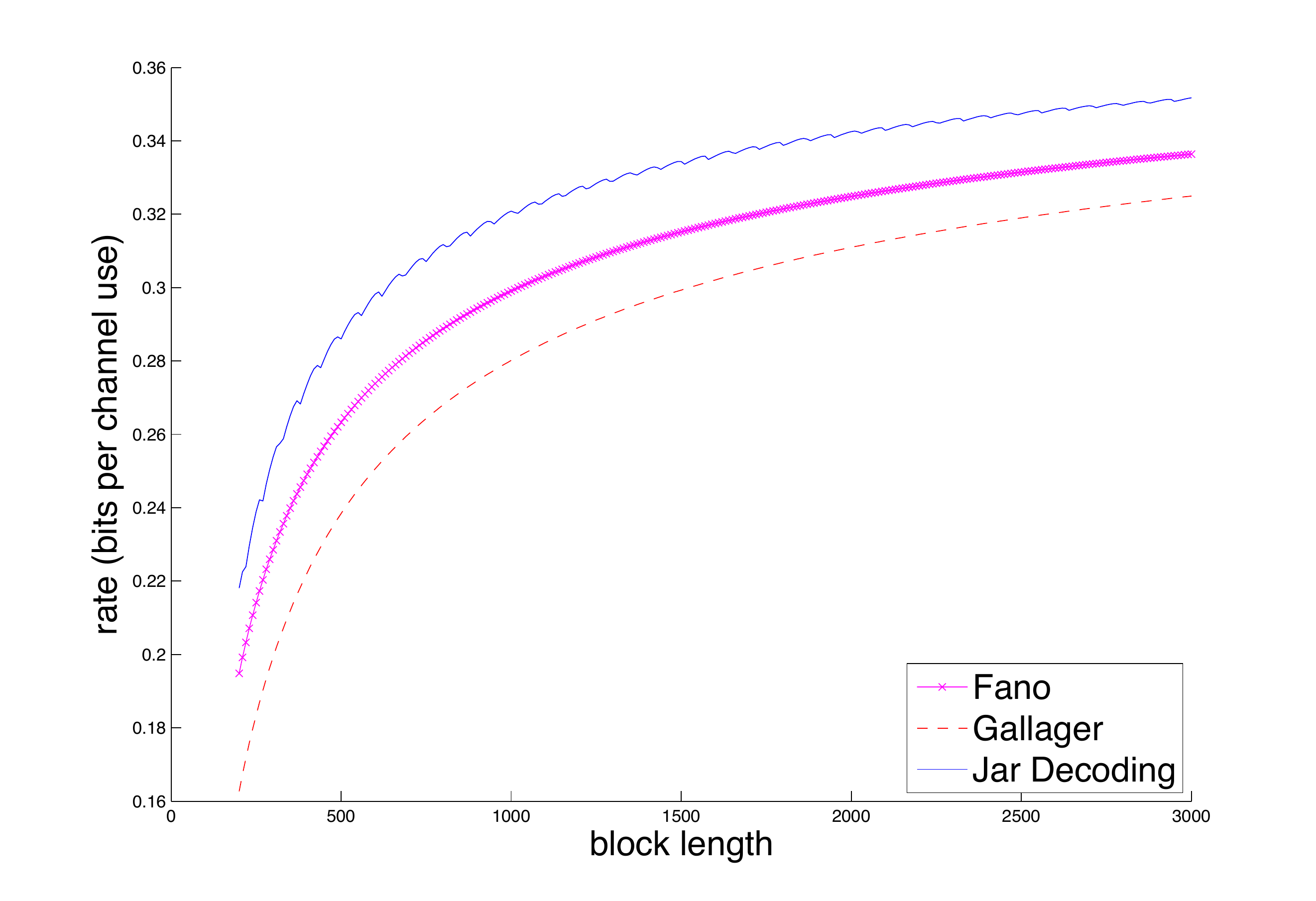}}
  \caption{Comparison of Achievability for Z Channel with $p=0.5$ and $P_e = 10^{-3}$}
  \label{fig:ach-z-1}
\end{figure}
\begin{figure}[h!]
  \centering
  \subfloat[$t = p_X$]{\includegraphics[scale=0.3]{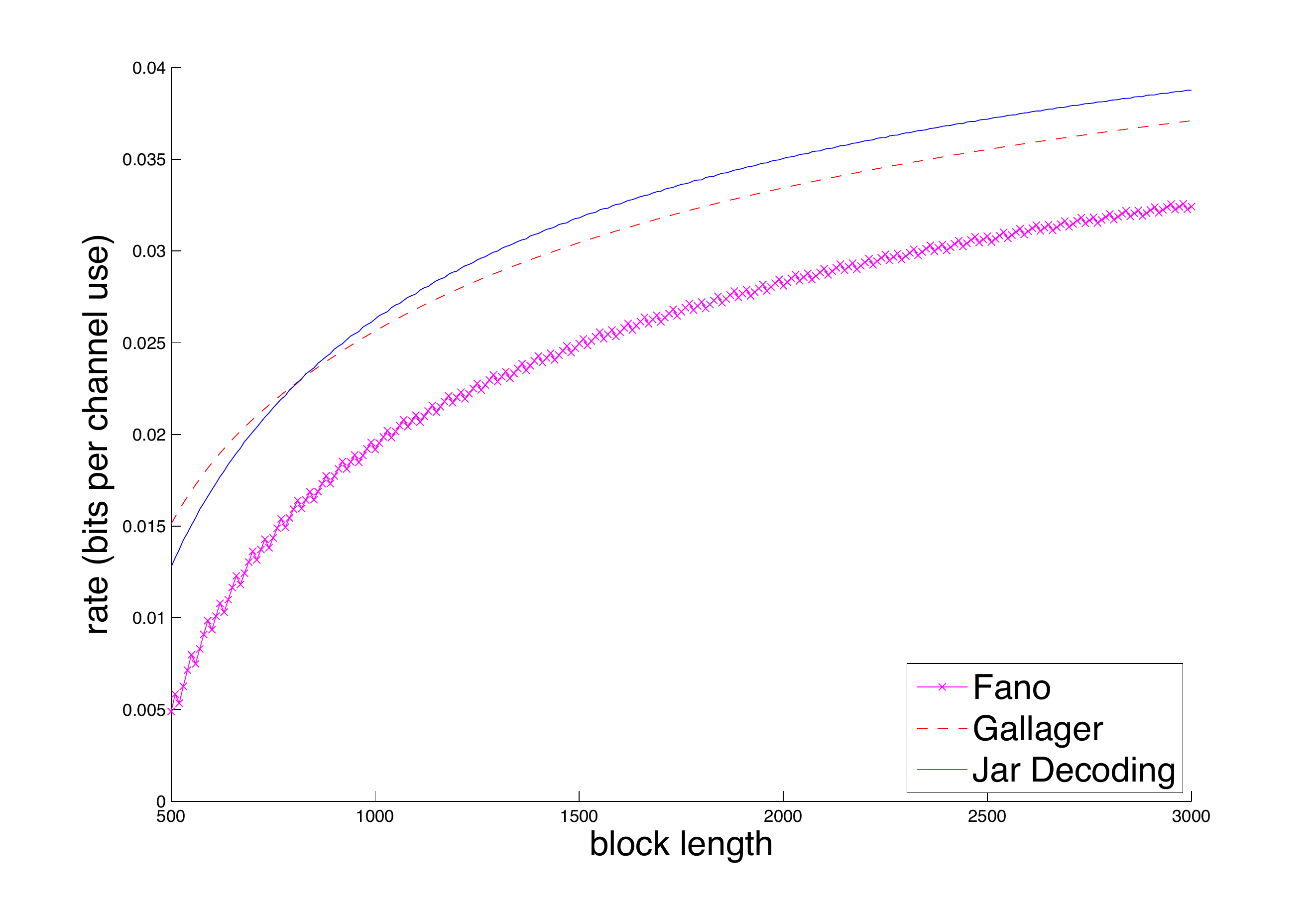}}
  \subfloat[$t = t^*$]{\includegraphics[scale=0.3]{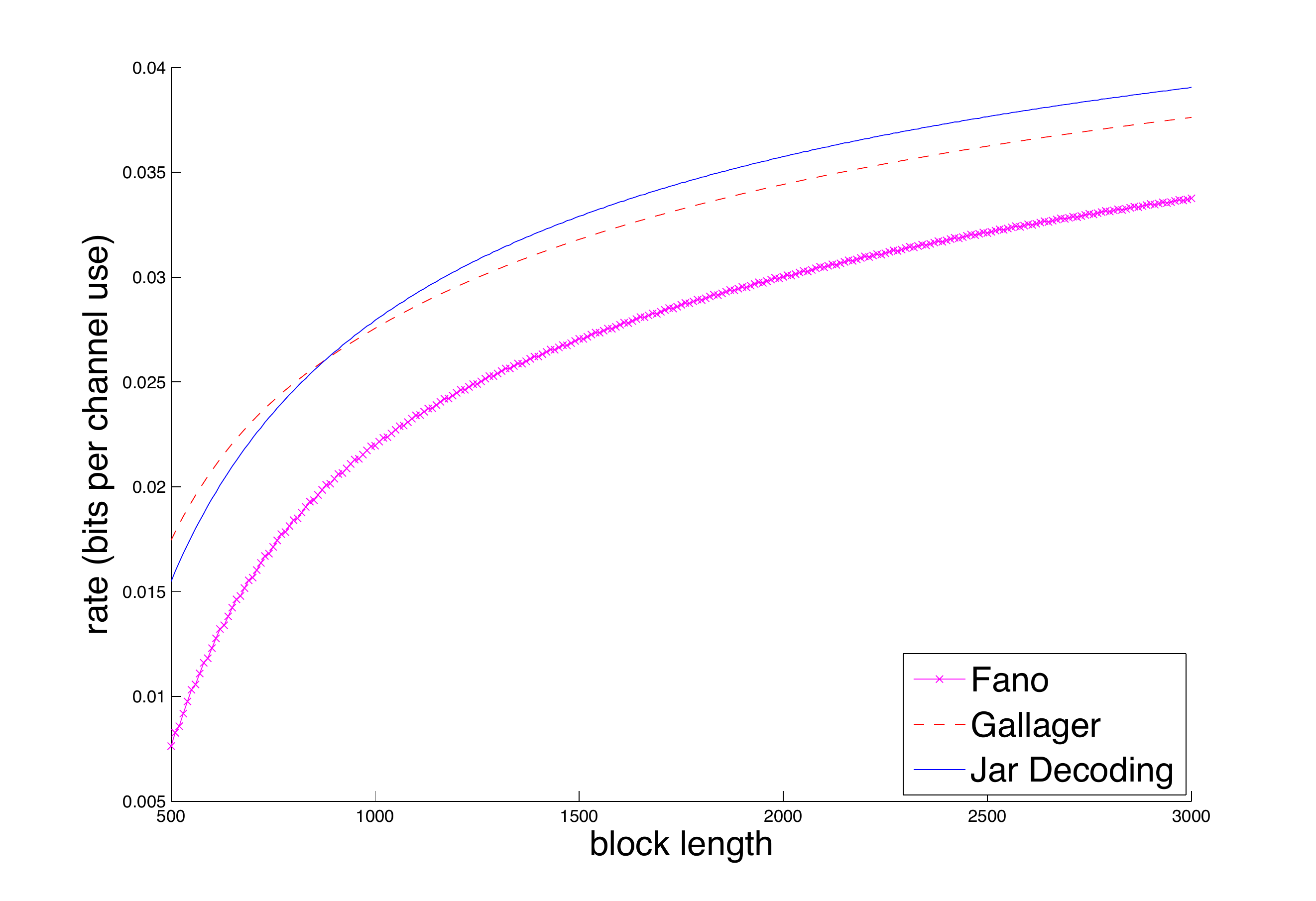}}
  \caption{Comparison of Achievability for Z Channel with $p=0.9$ and $P_e = 10^{-3}$}
  \label{fig:ach-z-2}
\end{figure}
As shown in Figures \ref{fig:ach-z-1} and \ref{fig:ach-z-2}, Theorem \ref{thm-random-dimc-1} constantly outperforms Fano's error exponent result. In addition, Figure \ref{fig:ach-z-1} shows that due to the non-exponential term $\left[ 1 + e^{n H(t)} |\mathcal{T}^n_t|^{-1} \right]$ or $e^{n H(t)} |\mathcal{T}^n_t|^{-1}$\footnote{In \cite{fano1961}, $e^{n H(t)} |\mathcal{T}^n_t|^{-1}$ is further bounded by $(2 \pi n)^{|X|} e^{|\mathcal{X}|/12}$.}, Fano's result could be worse than Gallager's, despite the relation of Fano's and Gallager's error exponent functions. Meanwhile, in Figure \ref{fig:ach-z-2}, $p_X$ represents the capacity achieving type, while $t^*$ is some type calculated in a way specified in \cite{jar-converse}. A close look at Figure \ref{fig:ach-z-2} then reveals that curves in (b) are above their counterparts in (a), which suggests that a capacity achieving input type or distribution is not necessarily optimal in the non-asymptotic regime.

\section{Conclusion}
\label{sec8}
\setcounter{equation}{0}

New non-asymptotic achievability bounds for random structured code ensembles, specifically Gallager parity check ensemble and Shannon random code ensemble with a fixed codeword type, have been derived for discrete input arbitrary output channels. These bounds are asymptotically tight up to the second order of the coding rate as the block length $n$ goes to infinity with either constant or sub-exponentially decreasing error probability $\epsilon$. When combined with non-asymptotic equipartition property (NEP) developed in this paper, they are also easy to compute for any discrete input arbitrary output channel. Numeric evaluation has demonstrated that our achievability bound on Gallager parity check ensemble is the tightest achievability result known so far in some non-asymptotic regime for binary input additive Gaussian channels. A key step in establishing these new bounds is the introduction of a decoding rule called jar decoding, which has led us to apply the union bound with respect to sequences inside a jar, instead of all codewords inside a codebook. The concept of jar decoding and its related bounding techniques, along with NEP, may be useful to non-asymptotical analysis of other problems in information theory as well. 

\appendices
\renewcommand{\theequation}{\Alph{section}.\arabic{equation}}
\setcounter{section}{0} \setcounter{equation}{0} %

\section{Non-asymptotic Equipartition Property with respect to conditional entropy}
\label{sec:non-aasympt-equip-cond}

In this appendix, we establish tight upper and lower bounds on $P_\delta$. In light of the asymptotic equipartition property (AEP) in the sense of the convergence of $- {1 \over n} \ln p(X^n |Y^n )$ to $H(X|Y)$ as $n \to \infty$ in probability, these bounds (i.e., in  \eqref{eq-nep-1}) will be referred to, with a slight abuse of the term ``equipartition'', as the non-asymptotic equipartition property (NEP) with respect to conditional entropy.

\begin{theorem}[NEP With Respect to $H(X|Y)$] \label{th-nep-cond}
For any positive integer $n$,
\begin{equation} \label{eq-wnep-1}
  \Pr \left \{ - {1 \over n} \ln p(X^n |Y^n ) > H(X|Y) + \delta \right \} \leq e^{- n r_{X|Y} (\delta) }
\end{equation}
where $X^n = X_1 X_2 \cdots X_n$,  $Y^n = Y_1 Y_2 \cdots Y_n$, and $(X_i, Y_i)$, $i=1, 2, \cdots, n$, are independent and identically distributed with $p(x, y)$ . Moreover,  under the assumptions \eqref{eq4-1} and \eqref{eq4-1-}, the following also hold:
 \begin{description}

 \item[(a)] There exists a $\delta^* >0$ such that for any $\delta \in (0, \delta^*]$,
 \begin{equation} \label{eq-nep-a-1}
    r_{X|Y} (\delta) = {1 \over 2 \sigma^2_H (X|Y) } \delta^2 + O(\delta^3)
  \end{equation}

  \item[(b)] For any $\delta \in (0, \Delta^* (X|Y))$ and any positive integer $n$
  \begin{eqnarray} \label{eq-nep-1}
  \underline{\xi}_{H}(X|Y,\lambda,n)  e^{- n r_{X|Y} (\delta)} &\leq&
  \Pr \left \{ - {1 \over n} \ln {p(Y^n |X^n)} > H(X|Y) + \delta \right \} \nonumber\\
  &\leq& \bar{\xi}_{H}(X|Y,\lambda,n)  e^{- n r_{X|Y} (\delta)}
  \end{eqnarray}
  where $\lambda = r'_{X|Y} (\delta) >0$, $ \bar{\xi}_{H}(X|Y,\lambda,n)$ is defined in \eqref{eq-nep-2-},  and
  \begin{equation}
    \label{eq-nep-3}
    \underline{\xi}_{H}(X|Y,\lambda,n) =
       e^{\frac{n \lambda^2 \sigma^2_{H}(X|Y,\lambda)}{2}} Q(\rho_*+\sqrt{n}  \lambda \sigma_{H}(X|Y,\lambda))
  \end{equation}
  with  $Q(\rho_*) = \frac{1}{2} - \frac{2CM_{H}(X|Y,\lambda)}{\sqrt{n} \sigma^3_{H}(X|Y,\lambda)}$.

 \item[(c)] For any  $ \delta \leq c \sqrt{\ln n \over n} $, where $c < \sigma_H (X|Y)$ is a constant,
  \begin{eqnarray} \label{eq-nep-4}
    Q  \left ( {\delta \sqrt{n} \over \sigma_H (X|Y)} \right ) - {C M_H (X|Y) \over \sqrt{n} \sigma^3_H (X|Y)}
    & \leq  &  \Pr \left \{ - {1 \over n} \ln p(X^n |Y^n ) > H(X|Y) + \delta \right \} \nonumber \\
  &  \leq &  Q  \left ( {\delta \sqrt{n} \over \sigma_H (X|Y)} \right ) + {C M_H (X|Y) \over \sqrt{n} \sigma^3_H (X|Y)} .
     \end{eqnarray}
  \end{description}
\end{theorem}

\begin{IEEEproof}
 The inequality \eqref{eq-wnep-1} follows from the Chernoff bound. To see this is indeed the case, note that
 \begin{eqnarray}
 \lefteqn{ \Pr \left \{ - {1 \over n} \ln p(X^n | Y^n ) > H(X|Y) + \delta \right \}
  } \nonumber \\
 & = & \Pr \left \{ -  \ln p(X^n | Y^n )  > n ( H(X|Y) + \delta)  \right \}
 \nonumber \\
 & \leq & \inf_{ \lambda \geq 0} { {\bf E} [ e^{- \lambda \ln p(X^n | Y^n)} ] \over e^{ n \lambda ( H(X|Y) + \delta)} } \nonumber \\
 & = &  \inf_{ \lambda \geq 0} e^{-n \left [ \lambda ( H(X|Y) + \delta) - \ln {\bf E} [ p^{-\lambda} (X_1|Y_1) ] \right ] } \nonumber \\
 & = &  \inf_{ \lambda \geq 0} e^{-n \left [ \lambda ( H(X|Y) + \delta) - \ln \iint p(y) p^{-\lambda +1} (x|y) d x dy  \right ] } \nonumber \\
 & = & e^{-n r_{X|Y} (\delta) }\;. \label{eq-wnep-proof-1}
 \end{eqnarray}
To show \eqref{eq-nep-a-1}, 
we first analyze the property of $ r_{X|Y} (\delta) $ as a function of $\delta$ over the region $\delta \geq 0$. Using a similar argument as in \cite[Properties 1 to 3]{yang-redundancy}, it is not hard to show that under the assumption \eqref{eq4-1}, $\delta(\lambda)$ as a function of $\lambda$ is continuously differentiable up to any order over $\lambda \in (0, \lambda^* (X|Y))$. Taking the first order derivative of $\delta (\lambda)$ yields
   \begin{eqnarray} \label{eq-wnep-proof-3}
     \delta' (\lambda) &=&   \iint {p(y) p^{-\lambda +1} (x|y) \over \left [ \iint p(v) p^{-\lambda+1} (u|v) d u dv \right ] } \left [ - \ln p(x|y) \right ]^2 d x dy  \nonumber \\
     &&{-}\: \left [ \iint { p(y) p^{-\lambda +1} (x) \over \left [ \iint p(v) p^{-\lambda +1} (u|v) d u dv \right ] } \left [ - \ln p(x|y) \right ] d x dy  \right ]^2     \nonumber \\
     & > & 0
       \end{eqnarray}
 where the last inequality is due to \eqref{eq4-1-}. It is also easy to see that $\delta (0) =0$ and $\delta'(0) = \sigma^2_H (X|Y)$. Therefore, $\delta (\lambda)$ is strictly increasing over $\lambda \in [0, \lambda^* (X|Y))$. On the other hand, it is not hard to verify that under the assumption \eqref{eq4-1}, the function $\lambda (H(X|Y) + \delta) -\ln \iint p(y) p^{-\lambda +1} (x|y) d x d y$ as a function of $\lambda$ is continuously differentiable over $\lambda \in [0, \lambda^* (X|Y))$ with its derivative equal to
  \begin{equation} \label{eq-wnep-proof-4}
   \delta - \delta(\lambda) \;.
  \end{equation}
  To continue, we distinguish between two cases: (1) $\lambda^* (X|Y) = \infty$, and (2) $\lambda^* (X|Y) < \infty$. In case (1), since $\delta (\lambda)$ is strictly increasing over $\lambda \in [0, \infty)$, it follows that for any $\delta = \delta (\lambda)$ for some $\lambda \in [0, \lambda^* (X|Y))$, the supremum in the definition of $r_{X|Y} (\delta)$ is actually achieved at that particular $\lambda$, i.e.,
 \begin{equation} \label{eq-wnep-proof-5}
 r_{X|Y} (\delta(\lambda)) = \lambda (H(X|Y) + \delta(\lambda) ) -
 \ln \iint p(y) p^{-\lambda +1} (x|y) d x dy \;.
  \end{equation}
  In case (2), we have that for any $\delta = \delta (\lambda)$ for some $\lambda \in [0, \lambda^* (X|Y))$ ,
   \begin{equation} \label{eq-wnep-proof-6}
 \beta (H(X|Y) + \delta(\lambda) ) - \ln \iint p(y) p^{-\beta +1} (x|y) d x dy  < \lambda (H(X|Y) + \delta(\lambda) ) - \ln \iint p(y) p^{-\lambda +1} (x|y) d x d y
  \end{equation}
 for any $\beta \in [0, \lambda^* (X|Y))$ with $\beta \not = \lambda$. In view of the definition of $\lambda^* (X|Y)$, \eqref{eq-wnep-proof-6} remains valid for any $\beta > \lambda^* (X|Y)$ since then the left side of \eqref{eq-wnep-proof-6} is $-\infty$. What remains to check is when $\beta = \lambda^* (X|Y)$. If
\[ \iint p(y) p^{-\lambda^* (X|Y) +1} (x|y)  d x d y= \infty \]
it is easy to see that \eqref{eq-wnep-proof-6} holds as well when $\beta = \lambda^* (X|Y)$.  Suppose now
\[ \iint p(y) p^{-\lambda^* (X|Y) +1} (x|y)  d x d y < \infty  \;.\]
In this case, it follows from the dominated convergence theorem that
 \[ \lim_{\beta \uparrow \lambda^{*} (X|Y) }  \iint  p(y) p^{-\beta +1}(x|y) d x d y =
     \iint p(y) p^{-\lambda^* (X|Y) +1}(x|y) d x d y \]
  and hence by letting $\beta$ go to $\lambda^* (X|Y)$ from the left, we see that
  \eqref{eq-wnep-proof-6} holds as well when $\beta = \lambda^* (X|Y)$. Putting all cases together, we always have that for any $\delta = \delta (\lambda)$ for some $\lambda \in [0, \lambda^* (X|Y))$,
   \begin{equation} \label{eq-wnep-proof-7}
 r_{X|Y} (\delta(\lambda)) = \lambda (H(X|Y) + \delta(\lambda) ) -
 \ln \iint p(y) p^{-\lambda +1} (x|y) d x d y \;.
  \end{equation}

  Let
  \[ \Delta^* (X|Y) \defeq \lim_{\lambda \uparrow \lambda^* (X|Y)} \delta (\lambda) \;.\]
 Since both $\delta (\lambda)$ and $ \ln \iint p(y) p^{-\lambda +1} (x|y) d x d y$ are continuously differentiable with respect to $\lambda \in (0, \lambda^* (X|Y))$ up to any order, it follows from \eqref{eq-wnep-proof-7} that $r_{X|Y} (\delta)$ is also continuously differentiable with respect to $\delta \in (0, \Delta^* (X|Y))$ up to any order. (At $\delta =0$, $r_{X|Y} (\delta)$ is continuously differentiable up to at least the third order inclusive.) Taking the first and second order derivatives of $r_{X|Y} (\delta)$ with respect to $\delta$, we have
 \begin{eqnarray} \label{eq-wnep-proof-8}
r'_{X|Y} (\delta) & = &  {d r_{X|Y} (\delta) \over d \delta } \nonumber \\
 & = & {d r_{X|Y} (\delta(\lambda)) \over d \lambda }  {d \lambda \over d \delta } \nonumber \\
 & = & {d r_{X|Y} (\delta(\lambda)) \over d \lambda }  {1  \over \delta' (\lambda) } \nonumber \\
 & = & {1  \over \delta' (\lambda) }  \left [ H(X|Y) + \delta (\lambda) + \lambda \delta' (\lambda)  -  \iint { p(y) p^{-\lambda +1} (x|y) \over \left [ \iint p(v) p^{-\lambda +1} (u|v) d u d v \right ] } \left [ - \ln p(x|y) \right ] d x d y  \right ] \nonumber \\
 & = & \lambda
 \end{eqnarray}
 and
 \begin{eqnarray} \label{eq-wnep-proof-9}
r''_{X|Y} (\delta) & = & {d \lambda \over d \delta } \nonumber \\
    & = & { 1 \over \delta' (\lambda) }
  \end{eqnarray}
where $\delta = \delta (\lambda)$.  Therefore, $r_{X|Y} (\delta)$ is convex, strictly increasing, and continuously differentiable up to at least the third order (inclusive) over $\delta \in [0, \Delta^* (X|Y))$. Note that from \eqref{eq-wnep-proof-8} and \eqref{eq-wnep-proof-9}, we have $r'_{X|Y} (0) = 0 $ and $r''_{X|Y} (0) = 1/\sigma^2_H (X|Y)$. Expanding $r_{X|Y} (\delta)$ at $\delta =0$ by the Taylor expansion, we then have that there exists a $\delta^* >0$ such that
 \begin{equation} \label{eq-wnep-proof-10}
 r_{X|Y} (\delta) = {1 \over 2 \sigma^2_H (X|Y)} \delta^2 + O(\delta^3)
 \end{equation}
 for $\delta \in (0, \delta^*]$.  

Now towards proving parts b) and c) of this theorem, by (\ref{eq-wnep-proof-7}), it is not hard to verify that
  \begin{eqnarray} \label{eq-nep-proof-2}
    \lefteqn{ \Pr \left \{ - {1 \over n} \ln p(X^n | Y^n ) > H(X|Y) + \delta \right \}}
    \nonumber \\
    & = & \iint \limits _{ - {1 \over n} \ln p(x^n | y^n ) > H(X|Y) + \delta } p(x^n, y^n) d x^n dy^n \nonumber \\
    & = & \iint \limits _{ - {1 \over n} \ln p(x^n | y^n) > H(X|Y) + \delta } f^{-1}_{\lambda} (x^n,y^n) f_{\lambda} (x^n,y^n) p(x^n, y^n)  d x^n dy^n \nonumber \\
    & = & \iint \limits _{ - {1 \over n} \ln p(x^n | y^n) > H(X|Y) + \delta }
              e^{ -n  \left [ -{1 \over n} \lambda \ln p(x^n | y^n) - \ln \iint p(v) p^{-\lambda +1} (u|v) d u d v \right ] }
              f_{\lambda} (x^n,y^n) p(x^n, y^n)  d x^n d y^n \nonumber \\
    & = & \iint \limits _{ - {1 \over n} \ln p(x^n | y^n) > H(X|Y) + \delta}
              e^{-n  \left [ -{1 \over n} \lambda \ln p(x^n|y^n) - \lambda (H(X|Y)+\delta) + r_{X|Y} (\delta)  \right ] }
              f_{\lambda} (x^n,y^n) p(x^n, y^n) d x^n d y^n   \nonumber \\
    & = & e^{- n r_{X|Y} (\delta)} \iint \limits _{ - {1 \over n} \ln p(x^n | y^n) > H(X|Y) + \delta}
              e^{-n  \lambda \left [ -{1 \over n} \ln p(x^n | y^n) - (H(X|Y)+\delta)  \right ] }
              f_{\lambda} (x^n,y^n) p(x^n, y^n) d x^n dy^n \nonumber \\
    & = & e^{-n r_{X|Y} (\delta)} \iint \limits _{ - {1 \over n} \ln p(x^n | y^n) > H(X|Y) + \delta}
              e^{- \sqrt{n}  \lambda \sigma_H(X|Y,\lambda) \frac{- \ln p(x^n | y^n) - n(H(X|Y)+\delta)}{\sqrt{n} \sigma_H(X|Y,\lambda)} }
              f_{\lambda} (x^n,y^n) p(x^n, y^n) d x^n d y^n \nonumber \\
    & = & e^{-n r_{X|Y} (\delta)} \int\limits_{\rho > 0} \iint\limits_{\frac{- \ln p(x^n|y^n) - n(H(X|Y)+\delta)}{\sqrt{n} \sigma_H(X|Y,\lambda)} = \rho}
             e^{- \sqrt{n}  \lambda \sigma_H(X|Y,\lambda) \rho} f_{\lambda} (x^n,y^n) p(x^n,y^n) d x^n d \rho \nonumber \\
    & = & e^{-n r_{X|Y} (\delta)} \int\limits^{+\infty}_{0} e^{- \sqrt{n}  \lambda \sigma_H(X|Y,\lambda) \rho} d (1-\bar{F}_n(\rho)) \nonumber \\
    & = & e^{-n r_{X|Y} (\delta)} \left[ \bar{F}_n(0) - \int\limits^{+\infty}_{0}
             \sqrt{n}  \lambda \sigma_H(X|Y,\lambda)  e^{- \sqrt{n}  \lambda \sigma_H(X|Y,\lambda) \rho}
             \bar{F}_n(\rho) d \rho \right]
  \end{eqnarray}
where the last equality is due to integration by parts,
  \begin{eqnarray*}
    \bar{F}_n(\rho) &\defeq& \Pr \left\{ \frac{- \ln p( \tilde{X}^n | \tilde{Y}^n) - n(H(X|Y)+\delta)}{\sqrt{n} \sigma_H(X|Y,\lambda)} > \rho \right\} \\
                  &=& \Pr \left\{ \sum^n_{i=1} \frac{- \ln p(\tilde{X}_i | \tilde{Y}_i) - (H(X|Y)+\delta)}{\sqrt{n} \sigma_H(X|Y,\lambda)} > \rho \right\}
  \end{eqnarray*}
and $\{ (\tilde{X}_i, \tilde{Y}_i) \}^n_{i=1}$ are IID random variable pairs with pmf or pdf (as the case may be) $f_{\lambda}(x,y)p(x,y)$. Let
\begin{eqnarray}
  \label{eq-nep-proof-3}
  \xi_n & \defeq & \bar{F}_n(0) - \int\limits^{+\infty}_{0}
             \sqrt{n}  \lambda \sigma_H(X|Y,\lambda)  e^{- \sqrt{n}  \lambda \sigma_H(X|Y,\lambda) \rho}
             \bar{F}_n(\rho) d \rho \\
  \label{eq-nep-proof-4}
            & = & \int\limits^{+\infty}_{0}
             \sqrt{n}  \lambda \sigma_H(X|Y,\lambda)  e^{- \sqrt{n}  \lambda \sigma_H(X|Y,\lambda) \rho}
             [\bar{F}_n(0) - \bar{F}_n(\rho)] d \rho \; .
\end{eqnarray}
At this point, we invoke the following central limit theorem of Berry and Esseen\cite[Theorem 1.2]{hall}.
\begin{lemma} \label{le1}
 Let $V_1, V_2, \cdots$ be  independent real random variables with zero means
and finite third moments, and set
 \[ \sigma_n^2 = \sum_{i=1}^n \be V_i^2 . \]
Then there exists a universal constant $C < 1$ such that for any $n \geq 1$,
 \[ \sup_{-\infty < t < +\infty} \left| \Pr \left\{ \sum_{i=1}^n V_i > \sigma_n t \right\}
  - Q (t) \right| \leq C \sigma_n^{-3} \sum_{i=1}^n \be |V_i |^3 . \]
\end{lemma}
Towards evaluating $\xi_n$, we can bound $\bar{F}_n(\rho)$ in terms of $Q(\rho)$, by applying Lemma \ref{le1} to $\{ - \ln p(\tilde{X}_i | \tilde{Y}_i) - (H(X|Y) +\delta) \}_{i=1}^n$. Then for $\rho > 0$, we have
\begin{eqnarray}
  \label{eq-nep-proof-5}
  \bar{F}_n(0) & \leq & Q(0) + \frac{CM_H(X|Y,\lambda)}{\sqrt{n} \sigma^3_H(X|Y,\lambda)} \nonumber \\
                & = & \frac{1}{2} + \frac{CM_H(X|Y,\lambda)}{\sqrt{n} \sigma^3_H(X|Y,\lambda)} \\
  \label{eq-nep-proof-6}
  \bar{F}_n(\rho) & \geq & \left[ Q(\rho)-\frac{CM_H(X|Y,\lambda)}{\sqrt{n} \sigma^3_H(X|Y,\lambda)}\right]^+
\end{eqnarray}
and
\begin{eqnarray}
  \label{eq-nep-proof-7}
  \bar{F}_n(0) - \bar{F}_n(\rho)
    & \geq & \left[ Q(0)-\frac{CM_H(X|Y,\lambda)}{\sqrt{n} \sigma^3_H(X|Y,\lambda)}
                           - \left(Q(\rho)+\frac{CM_H(X|Y,\lambda)}{\sqrt{n} \sigma^3_H(X|Y,\lambda)}\right)
                  \right]^+ \nonumber\\
    & = & \left[ \frac{1}{2} - Q(\rho) - \frac{2 CM_H(X|Y,\lambda)}{\sqrt{n} \sigma^3_H(X|Y,\lambda)} \right]^+
\end{eqnarray}
where $[x]^+ = \max \{ x, 0\}$. Now plugging \eqref{eq-nep-proof-5} and \eqref{eq-nep-proof-6} into \eqref{eq-nep-proof-2} yields
\begin{eqnarray}
  \label{eq-nep-proof-8}
  \xi_n &\leq& \frac{1}{2} + \frac{CM_H(X|Y,\lambda)}{\sqrt{n} \sigma^3_H(X|Y,\lambda)} -
                       \int\limits^{+\infty}_{0}
                       \sqrt{n}  \lambda \sigma_H(X|Y,\lambda)  e^{- \sqrt{n}  \lambda \sigma_H(X|Y,\lambda) \rho}
                       \left[ Q(\rho)-\frac{CM_H(X|Y,\lambda)}{\sqrt{n} \sigma^3_H(X|Y,\lambda)}\right]^+ d \rho \nonumber \\
            &=&    \frac{1}{2} + \frac{CM_H(X|Y,\lambda)}{\sqrt{n} \sigma^3_H(X|Y,\lambda)} -
                       \int\limits^{\rho^*}_{0}
                       \sqrt{n}  \lambda \sigma_H(X|Y,\lambda)  e^{- \sqrt{n}  \lambda \sigma_H(X|Y,\lambda) \rho}
                       \left[ Q(\rho)-\frac{CM_H(X|Y,\lambda)}{\sqrt{n} \sigma^3_H(X|Y,\lambda)}\right] d\rho \nonumber \\
            &=&    \frac{1}{2} + \frac{CM_H(X|Y,\lambda)}{\sqrt{n} \sigma^3_H(X|Y,\lambda)} -
                       \int\limits^{\rho^*}_{0}
                       \left[ Q(\rho)-\frac{CM_H(X|Y,\lambda)}{\sqrt{n} \sigma^3_H(X|Y,\lambda)}\right]
                       d \left( - e^{- \sqrt{n}  \lambda \sigma_H(X|Y,\lambda) \rho} \right) \nonumber \\
            &=&    \frac{2CM_H(X|Y,\lambda)}{\sqrt{n} \sigma^3_H(X|Y,\lambda)} +
                       \int\limits^{\rho^*}_{0} \frac{1}{\sqrt{2 \pi}} e^{-\frac{\rho^2}{2}}
                       e^{- \sqrt{n}  \lambda \sigma_H(X|Y,\lambda) \rho}
                       d\rho \nonumber \\
            &=&    \frac{2CM_H(X|Y,\lambda)}{\sqrt{n} \sigma^3_H(X|Y,\lambda)} +
                       \int\limits^{\rho^*}_{0} \frac{1}{\sqrt{2 \pi}}
                       e^{-\frac{(\rho+\sqrt{n}  \lambda \sigma_H(X|Y,\lambda))^2}{2}+\frac{n \lambda^2 \sigma^2_H(X|Y,\lambda)}{2}}
                       d\rho \nonumber \\
            &=&    \frac{2CM_H(X|Y,\lambda)}{\sqrt{n} \sigma^3_H(X|Y,\lambda)}
                       + e^{\frac{n \lambda^2 \sigma^2_H(X|Y,\lambda)}{2}}
                      \left[ Q(\sqrt{n}  \lambda \sigma_H(X|Y,\lambda)) - Q(\rho^*+\sqrt{n}  \lambda \sigma_H(X|Y,\lambda))\right]
                      \nonumber \\
            &=&    \bar{\xi}_H (X|Y,\lambda,n)
\end{eqnarray}
where $Q(\rho^*) = \frac{CM_H(X|Y,\lambda)}{\sqrt{n} \sigma^3_H(X|Y,\lambda)}$, and meanwhile plugging \eqref{eq-nep-proof-7} into \eqref{eq-nep-proof-2} yields
\begin{eqnarray}
  \label{eq-nep-proof-9}
  \xi_n &\geq& \int\limits^{+\infty}_{0}
                        \sqrt{n}  \lambda \sigma_H(X|Y,\lambda)  e^{- \sqrt{n}  \lambda \sigma_H(X|Y,\lambda) \rho}
                        \left[ \frac{1}{2} - Q(\rho) - \frac{2 CM_H(X|Y,\lambda)}{\sqrt{n} \sigma^3_H(X|Y,\lambda)} \right]^+ d\rho \nonumber \\
            &=&     \int\limits^{+\infty}_{\rho_*}
                        \sqrt{n}  \lambda \sigma_H(X|Y,\lambda)  e^{- \sqrt{n}  \lambda \sigma_H(X|Y,\lambda) \rho}
                        \left[ \frac{1}{2} - Q(\rho) - \frac{2 CM_H(X|Y,\lambda)}{\sqrt{n} \sigma^3_H(X|Y,\lambda)} \right] d\rho \nonumber \\
            &=&     \int\limits^{+\infty}_{\rho_*}
                        \left[ \frac{1}{2} - Q(\rho) - \frac{2 CM_H(X|Y,\lambda)}{\sqrt{n} \sigma^3_H(X|Y,\lambda)} \right]
                        d\left( -e^{- \sqrt{n}  \lambda \sigma_H(X|Y,\lambda) \rho} \right) \nonumber \\
            &=&     \int\limits^{+\infty}_{\rho_*} \frac{1}{\sqrt{2 \pi}} e^{-\frac{\rho^2}{2}}  e^{- \sqrt{n}  \lambda \sigma_H(X|Y,\lambda) \rho} d\rho \nonumber \\
            &=&     e^{\frac{n \lambda^2 \sigma^2_H(X|Y,\lambda)}{2}} Q(\rho_*+\sqrt{n}  \lambda \sigma_H(X|Y,\lambda)) \nonumber \\
            &=&     \underline{\xi}_H (X|Y,\lambda,n)
\end{eqnarray}
where $Q(\rho_*) = \frac{1}{2} - \frac{2CM_H(X|Y,\lambda)}{\sqrt{n} \sigma^3_H(X|Y,\lambda)}$.
Combining \eqref{eq-nep-proof-2} with \eqref{eq-nep-proof-8} and \eqref{eq-nep-proof-9} completes the proof of part (b) of Theorem \ref{th-nep-cond}.

Applying Lemma \ref{le1} to the IID sequence $\{-\ln p(X_i|Y_i) - H(X|Y) \}^n_{i=1}$, we get \eqref{eq-nep-4}. This completes the proof of Theorem \ref{th-nep-cond}.
\end{IEEEproof}

\begin{proposition}
  \label{cor-nep-asym}
  When $\lambda = o(1)$ and $\lambda = \Omega(1/\sqrt{n})$ as $n \rightarrow +\infty$, we have
  \begin{equation} \label{eq-cor-nep-asym-1}
     e^{\frac{n \lambda^2 \sigma^2_H(X,\lambda)}{2}} Q(\sqrt{n}  \lambda \sigma_H(X,\lambda))
      = \Theta \left( \frac{1}{\sqrt{n} \lambda} \right) = \omega \left( \frac{1}{\sqrt{n}} \right)
  \end{equation}
  and
  \begin{eqnarray}
    \label{eq-cor-nep-asym-2}
    \bar{\xi}_H(X|Y,\lambda,n) &=& e^{\frac{n \lambda^2 \sigma^2_H(X|Y,\lambda)}{2}} Q(\sqrt{n}  \lambda \sigma_H(X|Y,\lambda)) \left( 1 + O (\lambda) \right) \\
    \label{eq-cor-nep-asym-3}
    \underline{\xi}_H(X,\lambda,n) &=& e^{\frac{n \lambda^2 \sigma^2_H(X|Y,\lambda)}{2}} Q(\sqrt{n}  \lambda \sigma_H(X|Y,\lambda)) \left( 1 - O (\lambda) \right).
  \end{eqnarray}
\end{proposition}

\begin{IEEEproof}
Note that $\lambda = r'_{X|Y} (\delta) = \Theta (\delta)$. When $\lambda=\Omega(1)$ with respect to $n$, it can be easily verified that $\bar{\xi}_H (X|Y,\lambda,n)$ and $\underline{\xi}_H (X|Y,\lambda,n)$ are both on the order of $\frac{1}{\sqrt{n}}$, by applying well-known inequality
  \begin{equation}
    \frac{1}{t+t^{-1}} \frac{1}{\sqrt{2 \pi}} e^{-\frac{t^2}{2}} \leq Q(t) \leq \frac{1}{t} \frac{1}{\sqrt{2 \pi}} e^{-\frac{t^2}{2}}.
  \end{equation}
  Meanwhile, on one hand, it is easy to see that
  \begin{equation}
    \bar{\xi}_H(X|Y,\lambda,n) \leq e^{\frac{n \lambda^2 \sigma^2_H(X|Y,\lambda)}{2}} Q(\sqrt{n}  \lambda \sigma_H(X|Y,\lambda)) + \frac{2CM_H(X|Y,\lambda)}{\sqrt{n} \sigma^3_H(X|Y,\lambda)}.
  \end{equation}
  On the other hand,
  \begin{eqnarray}
    \underline{\xi}_H(X,\lambda,n)
                             &=& e^{\frac{n \lambda^2 \sigma^2_H(X|Y,\lambda)}{2}} Q(\sqrt{n}  \lambda \sigma_H(X|Y,\lambda)) -
                                     e^{\frac{n \lambda^2 \sigma^2_H(X|Y,\lambda)}{2}}
                                     \int\limits^{\rho_*+\sqrt{n}\lambda\sigma_H(X|Y,\lambda)}_{\sqrt{n}\lambda\sigma_H(X|Y,\lambda)}
                                     \frac{1}{\sqrt{2 \pi}} e^{-\frac{\rho^2}{2}} d \rho \nonumber \\
                             &=& e^{\frac{n \lambda^2 \sigma^2_H(X|Y,\lambda)}{2}} Q(\sqrt{n}  \lambda \sigma_H(X|Y,\lambda)) -
                                     e^{\frac{n \lambda^2 \sigma^2_H(X|Y,\lambda)}{2}}
                                     \int\limits^{\rho_*}_{0}
                                     \frac{1}{\sqrt{2 \pi}} e^{-\frac{(\rho+\sqrt{n}\lambda\sigma_H(X|Y,\lambda))^2}{2}} d \rho \nonumber \\
                             &=&  e^{\frac{n \lambda^2 \sigma^2_H(X|Y,\lambda)}{2}} Q(\sqrt{n}  \lambda \sigma_H(X|Y,\lambda)) -
                                      \int\limits^{\rho_*}_{0}
                                      \frac{1}{\sqrt{2 \pi}} e^{-\frac{\rho^2+2 \rho \sqrt{n}\lambda\sigma_H(X|Y,\lambda)}{2}} d \rho \nonumber \\
                             &\geq&  e^{\frac{n \lambda^2 \sigma^2_H(X|Y,\lambda)}{2}} Q(\sqrt{n}  \lambda \sigma_H(X|Y,\lambda)) -
                                      \int\limits^{\rho_*}_{0}
                                      \frac{1}{\sqrt{2 \pi}} e^{-\frac{\rho^2}{2}} d \rho \nonumber \\
                             &=&  e^{\frac{n \lambda^2 \sigma^2_H(X|Y,\lambda)}{2}} Q(\sqrt{n}  \lambda \sigma_H(X|Y,\lambda)) -
                                      \frac{2CM_H(X|Y,\lambda)}{\sqrt{n} \sigma^3_H(X|Y,\lambda)}.
  \end{eqnarray}
  To further shed light on $\bar{\xi}_H(X|Y,\lambda,n) $ and $\underline{\xi}_H(X|Y,\lambda,n) $, we observe that
  \begin{equation}
     \frac{1}{\sqrt{2 \pi} \sqrt{n} \lambda \sigma_H(X|Y,\lambda)+\frac{1}{\sqrt{2 \pi} \sqrt{n} \lambda \sigma_H(X|Y,\lambda)}}
     \leq e^{\frac{n \lambda^2 \sigma^2_H(X|Y,\lambda)}{2}} Q(\sqrt{n}  \lambda \sigma_H(X|Y,\lambda))
     \leq \frac{1}{\sqrt{2 \pi} \sqrt{n} \lambda \sigma_H(X|Y,\lambda)} .
  \end{equation}
  And therefore, whenever $\lambda = o(1)$ and $\lambda = \omega(n^{-1})$,
  \begin{equation}
     e^{\frac{n \lambda^2 \sigma^2_H(X,\lambda)}{2}} Q(\sqrt{n}  \lambda \sigma_H(X,\lambda))
      = \Theta \left( \frac{1}{\sqrt{n} \lambda} \right) = \omega \left( \frac{1}{\sqrt{n}} \right)
  \end{equation}
  which further implies
  \begin{eqnarray}
    \bar{\xi}_H(X|Y,\lambda,n) &=& e^{\frac{n \lambda^2 \sigma^2_H(X|Y,\lambda)}{2}} Q(\sqrt{n}  \lambda \sigma_H(X|Y,\lambda))
    \left( 1 + o(1) \right) \\
    \underline{\xi}_H(X,\lambda,n) &=& e^{\frac{n \lambda^2 \sigma^2_H(X|Y,\lambda)}{2}} Q(\sqrt{n}  \lambda \sigma_H(X|Y,\lambda))
    \left( 1 - o(1) \right).
  \end{eqnarray}
\end{IEEEproof}

\section{Non-asymptotic Equipartition Property with respect to Relative Entropy}
\label{sec:non-aasympt-equip-rela}
\setcounter{equation}{0} %

In this appendix, we establish tight upper and lower bounds on $P_{t, \delta}$. Once again, in light of the AEP with respect to relative entropy, these bounds (i.e., in \eqref{eqrl3-17}) are referred to as the NEP with respect to relative entropy. 

\begin{theorem}[NEP With Respect to Relative Entropy] \label{th-nep-rela}
For any sequence $x^n=x_1 \cdots x_n$ from $\cal X$, let $t \in {\cal P}$ be the type of $x^n$, i.e., $n t(a)$, $a \in {\cal X}$,  is the number of times the symbol $a$ appears in $x^n$. Then
  \begin{equation} \label{eqrl3-3}
  \Pr \left \{ \left. {1 \over n} \ln {p(Y^n |X^n)  \over q_t (Y^n) }\leq I(t; P) - \delta \right |  X^n =x^n    \right \} \leq e^{- n r_{-} (t, \delta) }\;.
  \end{equation}
 Furthermore,  under the assumptions \eqref{eq5-2} and \eqref{eq5-3+},  the following also hold:
 \begin{description}

 \item[(a)] There exists a $\delta^* >0$ such that for any $\delta \in (0, \delta^*]$
  \begin{equation} \label{eqrl3-4-}
 r_{-} (t, \delta) = {1 \over 2 \sigma^2_D (t; P) } \delta^2 + O(\delta^3)
  \end{equation}

  \item[(b)] For any $\delta \in (0, \Delta^*_{-} (t))$
  \begin{eqnarray} \label{eqrl3-17}
  \underline{\xi}_{D,-}(t;P,\lambda,n)  e^{- n r_{-} (t, \delta)} &\leq&
  \Pr \left \{ \left. {1 \over n} \ln {p(Y^n |X^n)  \over q_t(Y^n) }\leq I(t; P) - \delta \right | X^n = x^n  \right \} \nonumber\\
  &\leq& \bar{\xi}_{D,-}(t;P,\lambda,n)  e^{- n r_{-} (t, \delta)}
  \end{eqnarray}
  where $\lambda = {\partial  r_{-} (t, \delta)  \over \partial \delta} >0$, $\bar{\xi}_{D,-}(t;P,\lambda,n)$ is defined in \eqref{eqrl3-17-1-},    and
  \begin{equation}
    \label{eqrl3-17-2}
     \underline{\xi}_{D,-}(t;P,\lambda,n) =
       e^{\frac{n \lambda^2 \sigma^2_{D,-}(t;P,\lambda)}{2}} Q(\rho_*+\sqrt{n}  \lambda \sigma_{D,-}(t;P,\lambda))
  \end{equation}
  with $Q(\rho_*) = \frac{1}{2} - \frac{2CM_{D,-}(t;P,\lambda)}{\sqrt{n} \sigma^3_{D,-}(t;P,\lambda)}$.

  \item[(c)] For any  $ \delta \leq c \sqrt{\ln n \over n} $, where $c < \sigma_D (t; P)$ is a constant,
  \begin{eqnarray} \label{eqrl3-17+}
    Q  \left ( {\delta \sqrt{n} \over \sigma_D (t; P)} \right ) - {C M_D (t; P) \over \sqrt{n} \sigma^3_D (t; P)}
    & \leq  &   \Pr \left \{ \left. {1 \over n} \ln {p(Y^n |X^n)  \over q_t(Y^n) }\leq I(t; P) - \delta \right | X^n = x^n  \right \}
    \nonumber \\
    & \leq &  Q  \left ( {\delta \sqrt{n} \over \sigma_D (t; P)} \right ) + {C M_D (t; P) \over \sqrt{n} \sigma^3_D (t; P)}
     \end{eqnarray}
     where $0 < C < 0.56 $  is the universal constant in the Berry-Esseen central limit theorem \cite{id-clt-constant}.
  \end{description}
\end{theorem}

\begin{IEEEproof}
The inequality \eqref{eqrl3-3} comes from the Chernoff bound. To see this is indeed the case, note that
  \begin{eqnarray} \label{eq3-d1}
  \lefteqn{ \Pr \left \{ \left. {1 \over n} \ln {p(Y^n |X^n)  \over q_t (Y^n) }\leq I(t; P) - \delta \right |  X^n =x^n    \right \}} \nonumber \\
  & \leq & \inf_{\lambda \geq 0} { \be \left [ \left (  \left.  {p(Y^n |X^n)  \over q_t (Y^n) } \right )^{-\lambda} \right | X^n = x^n \right ] \over e^{n \lambda ( \delta - I(t; P))} } \nonumber \\
  & = & \inf_{\lambda \geq 0} { \prod_{a \in {\cal X}}  \left [ \int p(y|a )  \left ({p(y |a)  \over q_t (y) } \right )^{-\lambda}  d y  \right ] ^{n t(a)} \over e^{n \lambda ( \delta - I(t; P))} } \nonumber \\
  & = & \inf_{\lambda \geq 0} \exp \left \{ -n \left [  \lambda ( \delta - I(t; P)) - \sum_{a \in {\cal X}} t(a) \ln   \int p(y|a )  \left ({p(y |a)  \over q_t (y) } \right )^{-\lambda}  d y  \right ] \right \} \nonumber \\
  & = & e^{-n r_{-} (t, \delta) }
  \end{eqnarray}
  which completes the proof of \eqref{eqrl3-3}.

  The equation \eqref{eqrl3-4-} follows from the Taylor expansion of $r_{-} (t, \delta)$ at $\delta =0$ and the fact that
  \[   {\partial^2  r_{-} (t, \delta)  \over \partial \delta^2}   = { 1 \over \sigma^2_D (t; P) }\]
 at $\delta =0$ . What remains is to prove \eqref{eqrl3-17} and \eqref{eqrl3-17+}. To this end, let
 \[ f_{-\lambda} (y^n |x^n) = \prod_{i=1}^n f_{-\lambda} (y_i | x_i ). \]
 With $\lambda = {\partial  r_{-} (t, \delta)  \over \partial \delta}$, it follows from
 \eqref{eq3rp1} that
 \[  r_{ -} (t, \delta ) =   \lambda ( \delta - I(t; P)) -
  \sum_{x \in {\cal X}} t(x) \ln  \int  p(y |x) \left [ {p (y|x) \over q_t (y) } \right ]^{-\lambda}  d y  \;.\]
  Then we have
  \begin{eqnarray} \label{eq3-d2}
  \lefteqn{\Pr \left \{ \left. {1 \over n} \ln {p(Y^n |X^n)  \over q_t(Y^n) }\leq I(t; P) - \delta \right | X^n =x^n  \right \}} \nonumber \\
  & = & \int\limits_{{1 \over n} \ln {p(y^n |x^n)  \over q_t(y^n) }\leq I(t; P) - \delta} p(y^n | x^n) d y^n \nonumber \\
  & = & \int\limits_{{1 \over n} \ln {p(y^n |x^n)  \over q_t(y^n) }\leq I(t; P) - \delta}
            f^{-1}_{-\lambda} (y^n |x^n) f_{-\lambda} (y^n | x^n)  p(y^n | x^n) d y^n \nonumber \\
  & = & \int\limits_{{1 \over n} \ln {p(y^n |x^n)  \over q_t(y^n) }\leq I(t; P) - \delta}
            e^{ \lambda \ln {p(y^n |x^n)  \over q_t (y^n)} +
                  n \sum_{a \in {\cal X}} t(a) \ln   \int p(v|a )  \left ({p(v |a)  \over q_t (v) } \right )^{-\lambda}  d v  }
            f_{-\lambda} (y^n | x^n)  p(y^n | x^n) d y^n    \nonumber \\
  & = & \int\limits_{{1 \over n} \ln {p(y^n |x^n)  \over q_t(y^n) }\leq I(t; P) - \delta}
            e^{ \lambda \ln {p(y^n |x^n)  \over q_t (y^n)} +
                  n \lambda ( \delta - I(t; P)) - n r_{ -} (t, \delta )   }
            f_{-\lambda} (y^n | x^n)  p(y^n | x^n) d y^n    \nonumber \\
  & = & e^{-n r_{ -} (t, \delta )} \int\limits_{ \ln {p(y^n |x^n)  \over q_t(y^n) } - n(I(t; P) - \delta)  \leq 0}
            e^{ \lambda \left[ \ln {p(y^n |x^n)  \over q_t (y^n)} - n(I(t; P) - \delta) \right]   }
            f_{-\lambda} (y^n | x^n)  p(y^n | x^n) d y^n    \nonumber \\
  & = & e^{-n r_{ -} (t, \delta )} \int\limits_{\rho \leq 0}
            \int\limits_{ \frac{\ln {p(y^n |x^n)  \over q_t(y^n) }  - n(I(t; P) - \delta)}{\sqrt{n} \sigma_{D,-} (t;P,\lambda)} = \rho}
            e^{ \lambda \sqrt{n} \sigma_{D,-} (t;P,\lambda) \rho  }
            f_{-\lambda} (y^n | x^n)  p(y^n | x^n) d y^n    \nonumber \\
  & = & e^{-n r_{ -} (t, \delta )} \int\limits^{0}_{-\infty}
            e^{ \lambda \sqrt{n} \sigma_{D,-} (t;P,\lambda) \rho  } d F_{x^n} (\rho) \nonumber \\
  & = & e^{-n r_{ -} (t, \delta )} \left[ F_{x^n}(0) - \int\limits^{0}_{-\infty} \lambda \sqrt{n} \sigma_{D,-} (t;P,\lambda)
            e^{ \lambda \sqrt{n} \sigma_{D,-} (t;P,\lambda) \rho  } F_{x^n} (\rho) d \rho \right] \;.
   \end{eqnarray}
  where
  \begin{displaymath}
    F_{x^n} (\rho) =
    \Pr \left\{ \frac{\ln {p(Z^n |x^n)  \over q_t(Z^n) }  - n(I(t; P) - \delta)  }{\sqrt{n} \sigma_{D,-} (t;P,\lambda)} \leq \rho
                   \right\}
  \end{displaymath}
  and $Z_i$ takes values over the alphabet of $Y$ according to the pmf or pdf (as the case may be) $f_{-\lambda} (z | x_i)  p(z | x_i)  $.
  It is easy to verify that
  \[ \be \left [ \ln {p(Z_i |x_i)  \over q_t (Z_i) }  \right ] = D (t, x_i, \lambda) \]
  and
  \begin{eqnarray*}
    \sum_{i =1}^n \be \left [ \ln {p(Z_i |x_i)  \over q_t (Z_i) }  \right ]
  & = & \sum^n_{i=1} D(t,x_i,\lambda) \\
 &  =  & n \sum_{x \in {\cal X}} t(x) D(t, x, \lambda) \\
 & = & n(I(t; P) - \delta)
 \end{eqnarray*}
 which further implies that
 \begin{displaymath}
   F_{x^n} (\rho) =
    \Pr \left\{ \frac{ \sum^n_{i=1} \left[ \ln {p(Z_i |x_i)  \over q_t(Z_i) } - D(t,x_i,\lambda) \right]  }
                 {\sqrt{n} \sigma_{D,-} (t;P,\lambda)} \leq \rho \right\} \;.
 \end{displaymath}
 Applying Lemma~\ref{le1} to the independent sequence
  \[ \left \{ \ln {p(Z_i |x_i)  \over q_t (Z_i) } - D (t, x_i, \lambda)  \right \}_{i=1}^n , \]
 the argument similar to that in the proof of Theorem~\ref{th-nep-cond} can then be used to establish \eqref{eqrl3-17}.

  Finally, consider another sequence of independent random variables $W_1, W_2, \cdots, W_n$, where $W_i$ takes values over the alphabet of $Y$ according to the pmf or pdf (as the case may be) $ p(w | x_i)  $.  Applying Lemma~\ref{le1} directly to
  \[ \left \{ \ln {p(W_i |x_i)  \over q_t (W_i) } - D (t, x_i)  \right \}_{i=1}^n \]
we then get   \eqref{eqrl3-17+}.  This completes the proof of Theorem~\ref{th-nep-rela}.
\end{IEEEproof}

\bibliographystyle{IEEEtran}
\bibliography{IEEEabrv,channel,yang,math}

\end{document}